\documentclass[12pt,a4paper]{article}
\usepackage[round]{natbib}
\usepackage{bm}
\usepackage{url}
\usepackage{amsthm}
\usepackage{amsmath}
\usepackage{amssymb}
\usepackage{appendix}
\usepackage{booktabs}
\usepackage{csquotes}
\usepackage{enumitem}
\usepackage{etoolbox}
\usepackage{multirow}
\usepackage{bbm}
\usepackage{tabularx}

\usepackage{subcaption}
\usepackage{float}
\usepackage{commath}

\usepackage[boxed]{algorithm2e}

\numberwithin{equation}{section}
\usepackage{siunitx}
\robustify\bfseries

\setlist[description]{font=\normalfont\itshape\space}

\usepackage{bm} 
\usepackage{tikz}
\usetikzlibrary{positioning, fit, arrows.meta, shapes}
\usetikzlibrary{fit}
\usetikzlibrary{calc}
\usetikzlibrary{arrows}

\usepackage{tikz}
\usetikzlibrary{matrix}
\usetikzlibrary{shapes.callouts}
\usepackage{colortbl}

\definecolor{yellow}{rgb}{0.9,0.95,0}
\definecolor{green}{rgb}{0.2,0.8,0}

\tikzstyle{lobmatrix}=[matrix of nodes,%
    draw=white,%
    nodes={anchor=north,draw=black,inner sep=0pt,minimum height=0.5cm},%
    column sep=1ex,row sep=0.6ex,inner sep=2ex,%
    rounded corners,%
    column 1/.style={minimum width=1.5cm, nodes={}},%
    column 2/.style={minimum width=2.4cm},%
    column 3/.style={minimum width=1.5cm, nodes={}},%
    column 4/.style={minimum width=2.4cm, nodes={}},%
    f/.style={fill=blue!50}, e/.style={fill= red!50},
    ampersand replacement=\&
]

\definecolor[named]{AskRed}{RGB}{224,75,58}
\definecolor[named]{BidBlue}{RGB}{57,139,187}

\DeclareMathOperator*{\argmax}{arg\,max}

\DeclareMathOperator{\Inventory}{\mathrm{Inv}}

\usepackage{todonotes}

\title{Deep Hawkes Process for High-Frequency Market Making}

\author{Pankaj Kumar\thanks{Copenhagen Business School, Denmark. Email:pk.mpp@cbs.dk}} 

\date{}

\begin{document}
\maketitle

%
\section*{Abstract}

High-frequency market making is a liquidity-providing trading strategy that simultaneously generates many bids and asks for a security at
ultra-low latency while maintaining a relatively neutral position. The
strategy makes a profit from the bid-ask spread for every buy and sell
transaction, against the risk of adverse selection, uncertain execution
and inventory risk. We design realistic simulations of limit order
markets and develop a high-frequency market making strategy in which
agents process order book information to post the optimal price, order
type and execution time. By introducing the Deep Hawkes process
to the high-frequency market making strategy, we allow a feedback loop to be created between order arrival and the state of the limit order book,
together with self- and cross-excitation effects. Our high-frequency market making strategy accounts for the cancellation of orders that influence order queue position, profitability, bid-ask spread and the value of the order. The experimental results show that our trading agent outperforms the baseline strategy, which uses a probability density estimate of the fundamental price. We investigate the effect of cancellations on market quality and the agent's profitability. We validate how closely the simulation framework approximates reality by reproducing stylised facts from the empirical analysis of the simulated order book data.

%
\section{Introduction}
Technological innovations and regulatory initiatives in the financial
market have led to the traditional exchange floor being displaced by the
electronic exchange. The electronic exchange is a fully automated
trading system programmed to incisively enforce order precedence,
pricing and the matching of buy and sell orders. Each order's pricing,
submission and execution is performed using sophisticated algorithmic
trading strategies, which account for 85\% of the equity market's trading
volume \citep{Mukerji2019}. High-frequency trading (HFT, or high-frequency trader), a subset of algorithmic trading, is characterised by exceptionally high speeds, minuscule timeframes and complex programs for initiating and liquidating positions \citep{SEC2014}. The critical discussion on the role of HFT in a fragmented market has been reignited after the Flash Crash of 6 May 2010 \citep{Kirilenko2017}. This systemic intra-day anomaly only lasted for a couple of minutes, but temporarily wiped away a trillion dollars in market value. The analysis of agents resolved transaction level data in the E-mini by \cite{Kirilenko2017} also looks at the behaviour of market makers, whose inventory dynamics remain stationary in conditions of fluctuating liquidity. Even though the market design of E-mini has no high-frequency market maker liability, unlike equity markets, this seminal paper\citep{Kirilenko2017} gave a boost to research aimed at understanding high-frequency market making or other liquidity-providing strategies in an algorithmic trading setting.

Market making is a liquidity-providing trading strategy that quotes numerous bids and asks for a security in anticipation of making a profit from a bid-ask spread, while maintaining a relatively neutral position \citep{Chakrqborty2011}. The high-frequency market making strategy can be characterised as subset of HFT that uses latency, at a scale of nanoseconds, to trade in a fragmented market \citep{Menkveld2013}. The growing literature reports that the market makers provide quality liquidity, improve market quality, contribute to price efficiency and
have a positive but moderate welfare effect \citep{Kirilenko2017, Brogaard2014, Menkveld2013}. However, there is another strand in the literature that argues that the quality of liquidity is deceptive. The orders are characterised as phantom liquidity, which quickly disappears before other market participants can access it. The optimal design of market making strategies is therefore an important question for practical applicability, market design and security exchange regulations.

The research on market making spans numerous disciplines, including finance \citep{Ho1981, Glosten1985, Hara1986, Avellaneda2008, Gueant2013, Cartea2014, Ait2017}, agent-based modeling \citep{Das2005, Preis2006, Wah2017, Chao2018}, and artificial intelligence \citep{Spooner2018, Ganesh2019, Kumar2020}. Inspired by seminal work of \cite{Ho1981} and its mathematical formulation \citep{Avellaneda2008}, the quintessential research in finance considers market making as a stochastic optimal control problem. In a simplistic setting, the market is modelled as a stochastic process, in which market makers try to maximise the expected utility of their profit and loss under inventory constraints \citep{Gueant2013}. In parallel to inventory-based models, \cite{Glosten1985} proposed information-based models, in which market makers face adverse selection risk emerging from informed traders. The unrealistic assumptions placed on market models to mathematically extract the market maker's asset pricing forces researchers to look beyond stochastic optimal control approaches.

Market making has also been extensively investigated in agent-based modelling (ABM, or agent-based model) literature \citep{Das2005, Wah2017}. The ABMs in market making evolved from zero intelligence to an intelligent variant by incorporating order book microstructure for order placement, execution and pricing policy. For example, \cite{Toke2011} reinforced the zero-intelligence market maker model with order arrival following mutually exciting Hawkes processes. The Hawkes process has
been exhaustively used in an empirical estimation and calibration of
market microstructure models deemed essential for designing optimal
market-making strategies \citep{Toke2011, Alan2018, Morariupatrichi2018}. In these models, the arrival rate of orders is not dependent on the state of the limit order book. However, the empirical results suggest the existence of feedback loop between order arrival and the state of the limit order book, together with self- and cross-excitation effects, for which current models fail to account \citep{Gonzalez2017, Morariupatrichi2018}. In addition, the Hawkes process constrains the parametric specification for conditional intensity, which
limits the model's eloquence. To tackle the parametric specification
problem, \cite{Mei2017} proposed the Neural Hawkes process, in which the Hawkes process is generalised by calculating the event intensities from the hidden state of a long short-term memory (LSTM). Despite the success of the Neural Hawkes process in natural language processing \citep{Mei2017}, the facile LSTM architecture might be inadequate when it comes to modelling noisy, asynchronous order book events.

In recent years, deep learning has made significant inroads into high-
frequency finance. The Convolutional Neural Network (CNN)
architecture and its variants were used to model price-formation
mechanisms using order book events as input \citep{Sirignano2019, Cont2019, Tashiro2019, Tsantekidis2017}. However, the CNN architectures are not sophisticated enough to capture self- and cross-excitation effects in the limit order book (LOB) \citep{Zhang2019}. The deep long-short term memory (DLSTM) architecture performs a hierarchical processing of complex order book events, and as such is able to capture the temporal structures of LOB \citep{Sagheer2019A, Sagheer2019B}. However, training the DLSTM model directly through stochastic gradient descents, initialised with random parameters, may have led to the backpropagation algorithm being trapped within multiple local minima \citep{Sagheer2019B, Vincent2010}. To circumvent the aforementioned limitations, the literature proposed an unsupervised pre-training of each layer and a stacking of many convolutional layers \citep{Vincent2010}. We use Stacking Denoising Autoencoders (SDAEs) together with DLSTM to resolve the random weight initialisation problem in base architecture \citep{Sagheer2019B}. In addition, SDAEs are quite effective at filtering out noisy order-level data at minuscule resolutions.

The exemplary predictive performance of deep-learning models has
encouraged researchers to augment order book data with agent-based
artificial market simulation, for the purpose of investigating algorithmic trading strategies \citep{Maeda2019}. The success of the model is dependent on the simulation framework of the financial market being close to realism. However, algorithmic trading research is still waiting for market simulators that could be used for developing, training, and testing algorithms in a manner similar to classic Atari 2600 games simulator \citep{Mnih2015}. In this paper, we develop realistic simulations of the financial market and use them to design a high-frequency market making agent using the Deep Hawkes process (DHP). The DHP models the streams of order book events by constructing a self-exciting multivariate Hawkes process and a limit order state process, which are coupled and interact with each other. Based on a long stream of high-frequency transaction-level order book data for the different events (e.g. buy, sell, cancel, etc), the high-frequency market makers use DHP
to accurately predict every held-out event.

\subsection{Contribution}

This paper is the first to incorporate DHP into the market making
strategy, which allows feedback loops between order arrival and the
state of the limit order book, together with self- and cross-excitation
effects. We extend the neurally self-modulating multivariate point
process \citep{Mei2017} to the deep framework by stacking SDAE with DLSTM, resulting in DLSTM-SDAE. The SDAE resolves the problem associated with weight initialisation, multiple local minima and ultra-noisy order book data that the stacked recurrent network fails to address. Our approach outperforms the NHP in predicting the next order type and its time. The gained predictive power helps agents to outperform the benchmark market making strategy, and uses a probability density estimate of the fundamental price. We outline our contribution below:
\begin{enumerate}
\item We designed a multi-asset simulation framework that is scalable and
can augment markets of substantial size. The framework is built on
realistic market architecture, interface kernels, a matching engine and
the Financial Information eXchange (FIX) protocol. 
\item We are first to introduce a feedback loop between order arrival and the state of the order book using DHP in the high-frequency market making setting.
\item We investigate the predictive and trading performance of the agents
with the benchmark.
\item We explore the effect of cancellation on order queue position, agent's profitability, bid-ask spread, and value of order relative to queue position, in order to verify the existing empirical findings \citep{Moallemi2017, Dahlstrm2018}.
\end{enumerate}

\subsection{Structure}

The paper is organized as follow. Section \ref{sec:DHP_B} presents the background to the limit order book, market making, Hawkes process, Neural Hawkes process, and long short-term memory. Section \ref{sec:DHP_LR} explores the research stream in market making strategies. Section \ref{sec:DHP_DHP} explains the novel deep Hawkes process. Section \ref{sec:DHP_MASF} illustrates the multi-agent simulation framework. Section \ref{sec:DHP_E} elaborates on the experimental configuration. Section \ref{sec:DHP_R} provides the results of the experiments. Section \ref{sec:DHP_C} presents our conclusions.


%
\section{Background}
\label{sec:DHP_B}
In this section, we first introduce the limit order book together with basic definitions. We then briefly present examples of market making
strategy, and review essential tools for designing and investigating the
classical market making model. We start with the Hawkes process, the
assumptions of which are violated in the financial markets. To check the
missing links of the Hawkes process, we describe the Neural Hawkes
process. Finally, we discuss the LSTM framework, which represents a
divergence from selecting a parametric form for the conditional intensity
in the Hawkes process variants.

\subsection{Limit Order Books} 

\begin{figure}[H]
\begin{center}
\centerline{\includegraphics[width=\columnwidth]{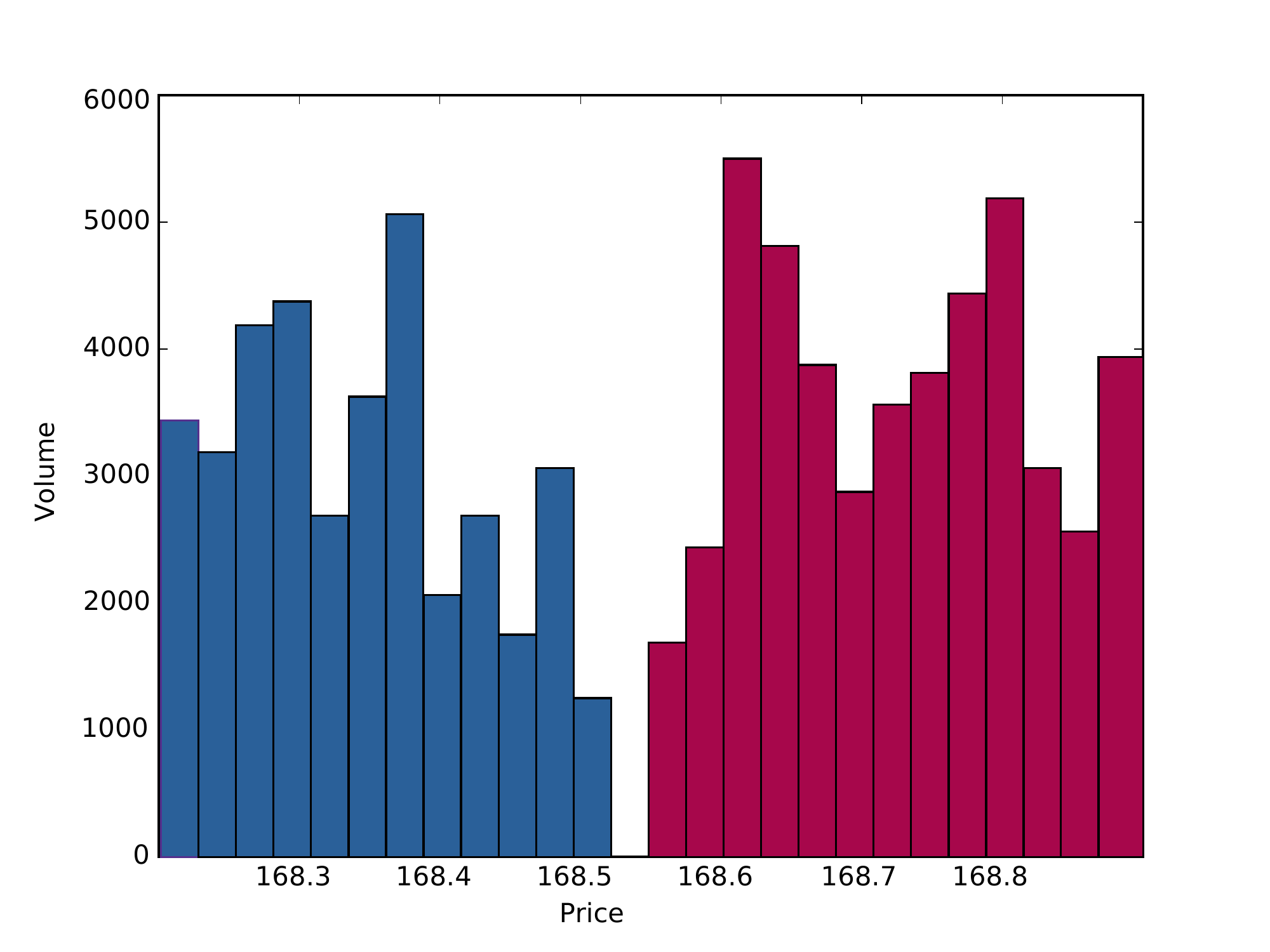}}
\caption{ Limit order book at NASDAQ for Apple Inc. (AAPL) on March 29th, 2018.}\label{fig:DHP_LOB}
\end{center}
\end{figure}

The LOB is a centralised database for outstanding orders submitted by traders to buy or sell a specified number of securities on an exchange.  Figure \ref{fig:DHP_LOB} illustrates an example of the reconstructed LOB for Apple securities traded on NASDAQ. The smallest increment by which the price of the security can move is called a \emph{tick}. The highest price at time $t$ for which there is outstanding buy order is called \emph{bid price}($168.60$), while the lowest sell price is called \emph{ask price} ($168.50$). The \emph{bid-ask spread} ($0.10$) at time $t$ is defined as the difference between the ask and bid prices. The \emph{mid price} (150.05) at time $t$ is the arithmetic average of the bid and the ask. For an in-depth review of definitions, mechanisms and
nomenclature, please refer to \cite{Gould2013}.

In an exchange, the traders can primarily submit three different types
of orders: \emph{limit orders}, \emph{market orders}, and \emph{cancellation orders}. A \emph{limit order} is an order to buy or sell a particular number of securities at a specified price or better. A \emph{market order} is an order to immediately buy or sell a certain number of securities at the best available price in the LOB. The arrival of market orders is instantly executed against the best available price in the limit order book. Orders that exceed the size available at the best price automatically spill over to the next best available price. Unlike market orders, the limit orders rest in the order book, pending
execution against a market order or being partially or fully \emph{canceled} by traders. The limit orders posted near to bid and ask are executed instantly, but may be prolonged if market prices diverge from the requested price. Market makers utilise this attribute to design optimal trading strategies \citep{Spooner2018}.  
 
In an exchange, several orders posted by traders can have the same
price at a given time $t$. To effectively match the orders within each
discrete price level, the LOBs employ various priority mechanism
algorithms. The algorithm most commonly employed by various
exchanges is \emph{price-time}. In this case, for buy or sell orders, the matching algorithms give priority to the orders with the highest or lowest price. In the event of ties, preference is given to orders with the earliest submission time relative to other orders \citep{Gould2013}. Other prominent priority mechanism algorithms include \emph{pro-rata} and \emph{price-size}. Under the \emph{pro-rata} algorithms, the ties at the given price are broken by distributing orders according to the depth in the LOB, while \emph{price-size} mechanism algorithms break the ties by giving higher priority to larger orders. Our study makes use of \emph{price-time} priority mechanism algorithms, as these encourage market makers to submit limit orders when designing their trading strategies. 

\subsection{High-Frequency Market Making}
Market making is a liquidity-providing trading strategy that quotes
numerous bids and asks for a security in anticipation of making a profit
in the form of bid-ask spread, while maintaining a relatively neutral
position \citep{Chakrqborty2011}. The modern market making strategy or high-frequency market making strategy can also be characterized as subset of HFT, where traders uses "lightning fast with a latency (inter-message time) upper bound of 1 millisecond, only engages in proprietary trading, generates many trades (it participates in 14.4\% of all trades, split almost evenly across both markets), and starts and ends most trading days with a zero net position" \citep{Menkveld2013}. The foundation of designing optimal market making strategies is dependent on optimising order-handling costs, adversely selected bid or ask quote costs, and non-zero position risk-averse costs\citep{Menkveld2013}.

In order to better understand high-frequency market making, let us consider a simple example from LOB illustrated in Figure \ref{fig:DHP_LOB}. The market making strategy places a buy order at $100.00$ and sell order at $100.10$. The execution of the both orders will give the traders profit of $00.10$, which is also the spread. As millions of market maker's trades are executed each second, they can amass huge profits.   

However, high-frequency market making strategies are always exposed to the risk of adverse price movements, uncertain executions and adverse
selections \citep{Penalva2015}. To avoid the above risks, the high-frequency market making strategies seek to compete in the market through lower order-processing costs, fast matching engines and low latency. The fast matching engine reduces the adverse selection risk by facilitating a
market-making strategy to immediately update quotes on the arrival
order book information, while low latency reduces the search for market
venues to mitigate a costly non-zero inventory position \citep{Penalva2015, Menkveld2013}. The array of sophisticated statistical or machine learning models used on the real-time order book data serve to
optimise the transaction costs, which are directly related to profitability.

\subsection{Hawkes Process}
To set the stage for Hawkes process, we first explicate key concept related to counting process, point process, and conditional intensity function from the classical literature \citep{Gao2018, Alan2018, Bacry2015, Embrechts2011}.

Let us consider a positive sequence of event arrival time, $\{t_n\}_{n \in \mathbb{N}}$, such that $\forall \ n \in \mathbb{N}$, $t_n < t_{n+1}$, defined on probability space $(\Omega,\mathcal{F},\mathbb{P})$ with almost surely finite, right-continuous step function defined for all $t \in \mathbb{R}_+$, and complete information filtration $(\mathcal{F}_{t})_{t\geq 0}$, such that $\mathbb{F} = \mathcal{F}^N(t)$, $t\geq 0$. Then the counting process $N(t)$ and its analogous point process $L(t)$ is defined as:

\begin{equation}
N(t)=\sum_{n \in \mathbb{N}} \mathbb{I}_{t_n \leq t} \quad \text{and} \quad L(t)=\sum_{n \in \mathbb{N}} \ell_{n}\cdot \mathbb{I}_{t_n \leq t}
\end{equation}
where $\mathbb{I}_{\{.\}}$ is the indicator function and $\{ \ell_{n}: n \ge 1 \}$ is a sequence of non-negative random variable having independent and identical distribution.

In academic literature, counting and point process terminology is often indistinguishable. The reader is expected to infer the nature of the
process depending on the context. For example, point process are often characterized by the distribution function of the
occurrence of the event conditioned to the past, but problems associated with the conditional arrival distribution mean that this is not very practical. Instead, the conditional intensity function is used. For counting process $N(t)$ with associated history $\mathcal{H}(t)$ and adapted to a filtration $\mathbb{F}$, we define the conditional intensity as: 

\begin{equation}\label{eq:CIF}
    \lambda(t|\mathbb{F}) = \lim_{h \rightarrow 0} \mathbb{E} \bigg[ \frac{(N(t+h) - N(t))|\mathcal{H}(t)}{h} \bigg| \mathbb{F} \bigg]
\end{equation}

We now define Hawkes process characterized by intensity $\lambda(t)$ with respect to its natural filtration as:

\begin{equation}\label{eq:HP}
    \lambda(t|\mathbb{F}) = \mu(t) + \int_{-\infty}^{t} \phi(t-\tau) \ {dN}(\tau)  
\end{equation}
where $\mu$ is an the baseline intensity and $\phi$ is a non-negative kernel function such that $||\phi||_1 = \int_0^{\infty}\phi(\tau)d\tau < 1$. Prominent kernel function  used in finance is exponential decay $\phi(t) = \alpha e^{-\beta t}$, where parameter $\alpha$ represents previous event weight and  $\beta$ past event duration. 

Now, we consider $N (t) = \{N_t^i\}_{i=1}^{M}$ as the M-dimensional  counting process, and its analogous point process as $\{L_t^i\}_{i=1}^{M}$. Similarly, we define the multidimensional Hawkes process as:
\begin{equation}\label{eq:MHP}
    \lambda_{i}(t|\mathbb{F})=\mu_i+\sum_{j=1}^{M}\int_0^t \phi_{i,j}(t-\tau)d N_j(\tau),
    \end{equation}
where, $\mu = \{\mu_i\}_{i=1}^{M}$ a baseline intensity vector and $\phi(t) =  \{\phi_{i,j}(t)\}_{i,j=1}^{M}$ a matrix-valued kernel that is component-wise non-negative, causal, and $L^1$-integrable \citep{Bacry2015}.

We can also enrich the Equation \ref{eq:MHP} by associating each event with time of event $t_n$, its component $\varkappa _n$ and mark $k_n$. For example, modeling the trades performed at event time $t_n$ with different volume $k_n$ and drawdown intensity $\varkappa_n$ \citep{Bacry2015}.  The vector intensity function of the multivariate marked Hawkes process will be defined as: 

\begin{equation}
    \lambda_{i}(t,n |\mathbb{F})=\mu_i+\sum_{j=1}^{M} \int_0^t \phi_{i,j}(t-\tau)\psi_{j}(n)d N_j(\tau \times \delta),
    \end{equation}
where, $\mu = \{\mu_i\}_{i=1}^{M}$ a exogenous intensities vector, $\phi(t) =  \{\phi_{i,j}(t)\}_{i,j=1}^{M}$ a matrix-valued kernel and $\psi_{j}(n)$ a impact function of marks.

The properties of the Hawkes process can be characterized thoroughly in an analytical manner due to the linear structure of the stochastic intensity \citep{Alan2018, Bacry2015}.The linear properties of a Hawkes process enable linear predictions of the models, given their base intensity and the kernels as parameters (the kernels are non-parametrically calibrated from the data). A Hawkes process can be appropriately approximated as auto-regressive process, wiener processes, and clustering representation \citep{Bacry2015}. Simply put by \cite{Mei2017}, " the Hawkes process supposes that past events can temporarily raise the probability of future events, assuming that such excitation is positive, additive over the past events, and exponentially decaying with time". In a simplified setting, these properties might be useful for modelling constrained processes, but might not be applicable to real world examples. For example, at large scales, the price formation process's microscopic variables, as they relate to order book events, do not diffuse toward Wiener processes. Similarly, the massive cancellation of orders in LOB might inhibit price rather than exciting it.

\subsection{Neural Hawkes Process} \label{ss:NHP}

The Neural Hawkes Process was (NHP) introduced to fill the gaps in the Hawkes process's unrealistic assumptions. Building on the earlier
formulation, the positivity constraints on baseline intensity vector $\mu$, kernels $\phi$, and decay rate $\delta$ limit the eloquentness of Hawkes process. It fails to capture inhibition and inherent inertia effect, which are characteristics of realistic financial market \citep{Mei2017}.

Let $\{t_n, \varkappa_n, k_n\}_{n \in \mathbb{N}}$ is a event streams, where $t_n$ are times of occurrence of an event, its component $\varkappa_n$, and their corresponding mark $\{k_n\}_{n \in \mathbb{N}}$ in $k_n:= \{1, \dots ,K\}$. Then, the probability of incidence of next event at time $t_n$ of type $k_n$ is $\mathbb{P}\{(t_n,\varkappa_n, k_n) \mid \mathcal{H}_n, (t_n - t_{n-1})\} dt$. The associated intensity function conditioned on the past events $\overline{h}$ for self-exciting multivariate point process or Hawkes process with exponentially decaying kernel function is:
\begin{equation}
    \lambda_k(t)= \mu_k + \sum_{\overline{h}: t_{\overline{h}} < T}= \alpha_{k_{\overline{h}},k} \exp (-\beta_{k_{\overline{h}},k} (t-t_{\overline{h}}) ),
    \end{equation}
The inhibition and inherent inertia effect are introduced in the self-modulating model, where we  relax positivity constraints on parameters $\alpha$ and $\mu$. The negative total resultant activation is then passed through non-linear transfer function (e.g. rectified linear unit (ReLU) function, softplus function etc.) such that:

\begin{equation}
    \lambda_k(t) = f_k(\tilde{\lambda}_k(t)),
    \end{equation}

\begin{equation} \label{eq:NHP}
    \tilde{\lambda}_k(t)= \mu_k + \sum_{\overline{h}: t_{\overline{h}} < t} \alpha_{k_{\overline{h}},k} \exp (-\beta_{k_{\overline{h}},k} (t-t_{\overline{h}}) ).
    \end{equation}
where $\mu_k < 0$, $\alpha_{j,k} < 0$ allows inertia and inhibition effect respectively \citep{Mei2017}.
    
The summation in Equations \ref{eq:NHP} places a restriction on the $\tilde{\lambda}_k(t)$, where past events have an independent and additive influence. This deviates from reality, which is characterised by the existence of complex dependence between the intensities in terms of number of order event types and past event timings \citep{Bacry2015}.  \cite{Mei2017} proposed the \emph{Neural Hawkes Process}  to learn and predict complex dependency by replacing the summation with a novel recurrent neural network. In this novel process, the hidden state vector $\mathbf{h}(t)$ controls the dynamics of time varying event intensities, which in turn depends on a vector $\mathbf{c}(t)$ of memory cells in a continuous-time long short-term memory \citep{Hochreiter1997}.
 
\subsection{Long Short-Term Memory} \label{ss:lstm}

\begin{figure}[H]
\begin{center}
\centerline{\includegraphics[width=\columnwidth]{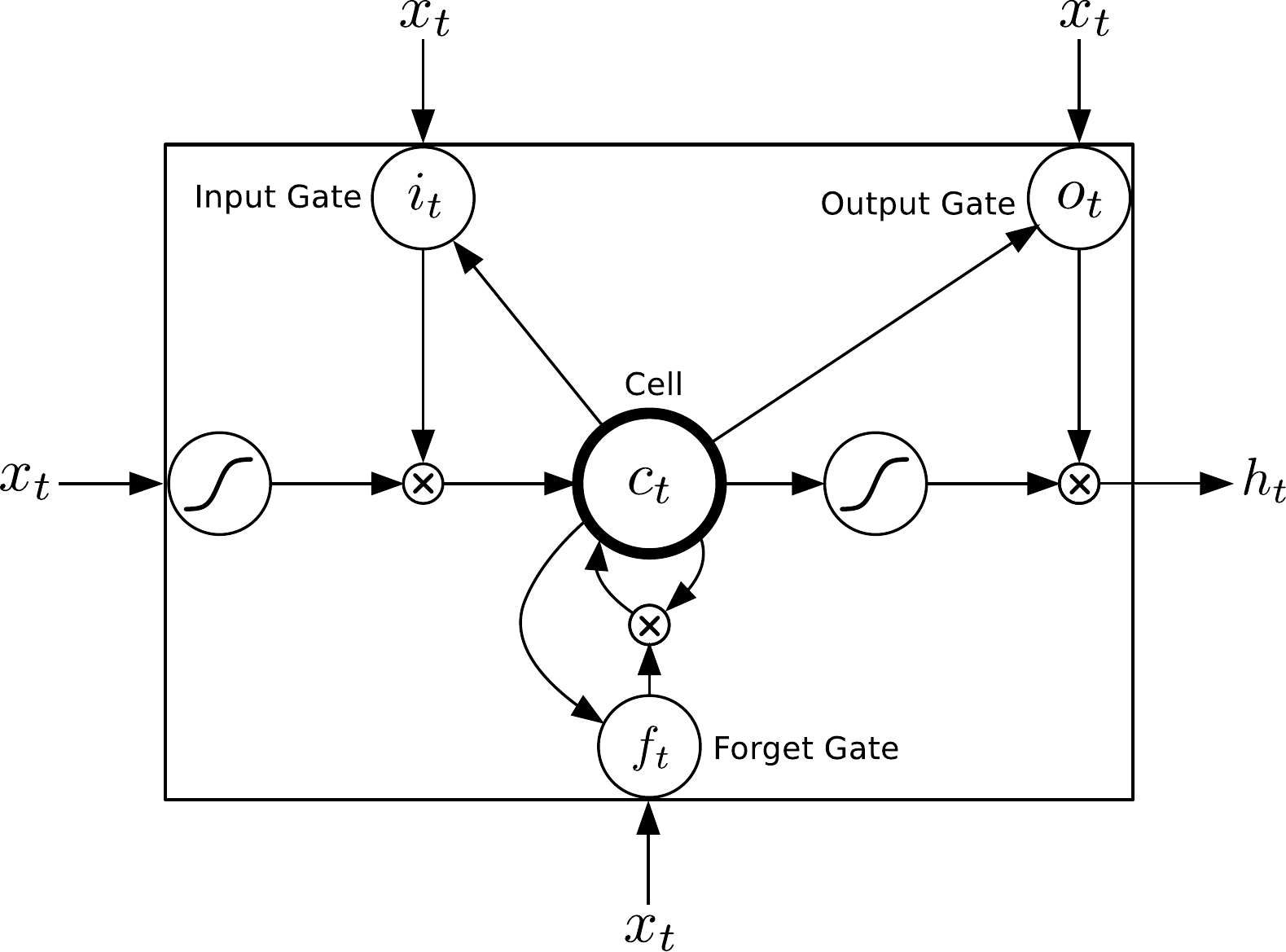}}
\caption{ LSTM Memory Cell \citep{Zhu2016}.}\label{fig:lstm}
\end{center}
\end{figure}

The central idea of LSTM is the use of {\em memory cell}, which overcomes the problem associated with {\em vanishing gradient} \citep{Arras2019}. The memory cell of LSTM is a complex unit, built from alike nodes in a distinct connectivity pattern, with the novel inclusion of multiplicative nodes, represented in figure \ref{fig:lstm}. A typical LSTM memory cell architecture contains a cell input activation vector ${{x}}_t$, an input gate ${{i}}_t$, a forget gate ${{f}}_t$, a cell ${{c}}_t$, an output gate ${{o}}_t$ and an output response $h_t$. The distinctive feature of the LSTM approach, input gate and forget gate, govern the information flow into and out of the cell according to gate logic. Whereas, the output gate controls the amount of information flow from the cell to the output ${{h}}_t$. A self-connected recurrent edge with fixed unit weight in the memory cell ensures that error can flow across many time steps without vanishing or exploding.

At each time step $t$, the recursive computation in the LSTM model proceeds according to the following equations:

\newcommand{\tanhmine}{\operatorname{tanh}}
\newcommand{\drop}{\mathring} 

\begin{eqnarray}
\label{eqn:lstm}
\begin{aligned}
\!&{\bf{i}}_t = \sigma \left( {{\bf{W}}_{xi}}{{\bf{x}}_{t}} + {{\bf{W}}_{hi}}{\bf{h}}_{t-1} + {{\bf{W}}_{ci}}{\bf{c}}_{t-1} + {\bf{b}}_i \right), \\ 
\!&{\bf{f}}_t = \sigma \left( {{\bf{W}}_{xf}}{{\bf{x}}_{t}} + {{\bf{W}}_{hf}}{\bf{h}}_{t-1} + {{\bf{W}}_{cf}}{\bf{c}}_{t-1} + {\bf{b}}_f \right), \\
\!&{\bf{\overline{c}}}_t = \phi \left( {{\bf{W}}_{xc}}{{\bf{x}}_{t}}\! +\! {{\bf{W}}_{hc}}{\bf{h}}_{t-1} \!+\! {\bf{b}}_c \right), \\ 
\!&{\bf{c}}_t = {\bf{f}}_t\!\odot {\bf{c}}_{t-1}\!+ {\bf{i}}_t\!\odot {\bf{\overline{c}}}_t, \\ 
\!&{\bf{o}}_t = \sigma \left( {{\bf{W}}_{xo}}{{\bf{x}}_{t}} + {{\bf{W}}_{ho}}{\bf{h}}_{t-1} + {{\bf{W}}_{co}}{\bf{c}}_{t} + {\bf{b}}_o \right), \\
\!&{\bf{h}}_t = {\bf{o}}_t \odot \phi \left( {\bf{c}}_t \right),
\end{aligned}
\end{eqnarray}
where ${{x}}_t$ is input vector at time $t$, the activation function $\sigma \left(x \right)$ / $\phi \left(x \right)$ is defined as sigmoid $\sigma\left(x\right)=1/(1+e^{-x})$ / $\phi \left(x \right)=\tanh$, ${\bf{W}}_{A B}$ is the weight matrix between $A$ and $B$ (e.g., ${\bf{W}}_{x i}$ is the weight matrix from the inputs ${\bf{x}}_{t}$ to the input gates ${\bf{i}}_t$), ${\bf{b}}_{B}$ denotes the bias term of $B$ with $B \in \{i,f,c,o\}$ and $\odot$ denotes point-wise multiplication of the two vectors. To ensure comprehensibility with standard literature, we follow
the same naming conventions discussed in the paper by \cite{Zhu2016}.


%
\section{Literature Review}
\label{sec:DHP_LR}

In this section, we briefly review three prominent streams of research
into how the Hawkes process is used to design market making strategies.
We are well aware of blurred and overlapping boundaries across
research streams, but strongly believe in capturing the essence of
Hawkes models in market making.

\subsection{Stochastic Optimal Control}

The classical finance literature employs a stochastic optimal control
framework to determine market maker's optimal quotes for bid and ask
in the presence of inventory risk, adverse selection, information
asymmetry and latency \citep{Ho1981, Sandas2001, Avellaneda2008, Penalva2015}. The market maker aims to maximise the expected utility of the profit and loss contour at closing time. By integrating a utility framework into the microstructure of LOB, \cite{Avellaneda2008} were able to analytically derive optimal bid and ask quotes. In the continuous-time model, the mid-prices evolve according to Brownian motion, and the order arrival follows a Poisson process. In one particular case, a non-homogeneous Poisson process is reduced to the Hawkes process \citep{Bacry2015}. The unrealistic assumptions limit the model’s ability to capture adverse selection effects, market impact and autocorrelation structures in the security being traded. Building on the seminal work off \cite{Avellaneda2008}, numerous researchers looked at price impact, adverse selection effects, and latency, together with different objective functions \citep{Penalva2015}. The purpose of all of the improvements in the model is to provide numerical approximations for the associated
stochastic differential equations. However, problems related to model
ambiguity have yet to be addressed \citep{Nystrom2014}. 

The profusion of transaction-level data at a minuscule resolution
provides an unprecedented opportunity to apply Hawkes processes to
the study of market microstructure with a view to deciphering price
formation mechanism, liquidity dynamics and volatility \citep{Morariupatrichi2018}. The use of Hawkes processes to model
extreme price moves and order flow is well integrated with the high-
frequency market making model \citep{Nystrom2014, Bacry2015}. For example, \cite{Nystrom2014} proposed a market making
model based on model risk or uncertainty, in which inventory dynamics,
fill rates and price formation are modelled using two independent
Hawkes processes. While the market making models based on the
Hawkes processes have been moderately successful in integrating a
realistic order arrival process, they fail to incorporate the complex
interaction between the self- and cross-excitation effects of the
endogenous state variables that describe the LOB \citep{Morariupatrichi2018}. In the classical "buy low and sell high" market making strategies, \citep{Cartea2014} used a multivariate Hawkes process to capture the interaction between market orders and the state of LOB. However, the constraints placed on parametric specification for
conditional intensity by the Hawkes process limit the model's
eloquence.
 
\subsection{Agent-Based Models}

Market making has been also been extensively investigated in ABM \citep{Das2005, Wah2017}. Using a bottom-up approach, the ABMs try to artificially simulate the systems' aggregate behaviours caused by the actions and interactions between heterogeneous autonomous agents \citep{Samanidou2007}. The ABMs in market making range from the simple "zero intelligence" of mainstream economics to representing in detail the full complexity of the order book and market microstructure. The seminal work of \cite{Toke2011} assimilated a microscopic, dynamical statistical model for the continuous double auction \citep{Smith2003} into  Hawkes framework. The reinforced zero-intelligence market making model
populated with the liquidity provider and liquidity taker uses mutually
exciting Hawkes processes for the order arrival process. However, the
parametric specification of the exponential kernels contradicts the
empirical results on the existence of a feedback loop between order
arrival and the state of LOB \citep{Gonzalez2017, Morariupatrichi2018}. The Neural Hawkes process \cite{Mei2017} applied a Neural
Hawkes process to natural language processing in order to tackle the
parametric specifications problem, and generalised the Hawkes process
by calculating the event intensities from the hidden state of an LSTM.

Another prominent ABM in market making research \citep{Wah2017}uses empirical game-theory analysis to examine the effect of
market making on market performance in different scenarios. In the
paper, the authors use multiple background traders (e.g. traditional zero
intelligence agents) to produce realistic market microstructure. The
Bayesian market makers were then used to investigate welfare effect,
trading gains and strategic behaviour. The simplistic adaptive trading
strategies adopted by the market makers ought to be sufficient to
investigate market equilibria, but would face serious challenges in
relation to developing realistic market making strategies that also take
account of market microstructure. The academic literature needs to
assimilate the success of artificial intelligence in designing trading
strategies that interact with close-to-reality market simulations.

\subsection{Artificial Intelligence}

The success of deep reinforcement learning in board games \citep{Silver2017} and video games \citep{Mnih2015} soon sent ripples through the world of finance. Drawing on the classical mathematical
setup by \cite{Avellaneda2008}, \cite{Gueant2019} used a model-based deep actor-critic algorithm to find the optimal bid and ask quotes across a high-dimensional space of corporate bonds. \cite{Spooner2018} rreconstructed a limit order book from historical data and used it to construct a market making agent using temporal-difference reinforcement learning. The next natural progression, a multi-agent simulation of a dealer market, was developed by \cite{Ganesh2019} to understand the behaviour of market making agents. The success of this model is dependent on a realistic simulation framework of the financial market. However, algorithmic trading research is evolving, and is still at the stage of exploiting market simulators that could be used to develop training and testing algorithms in a similar vein to the classic Atari 2600 games simulator \citep{Mnih2015}.

Despite the exemplary predictive performance of deep learning
models, market making has not yet been comprehensively addressed. If
we return to stochastic optimal control in market making, we notice that
the dynamics of order flow are mostly described by variants of Hawkes
processes, while the resultant control algorithms are associated with
continuous semi-martingales, which are hard to solve recursively \citep{Gueant2019}. In addition, the unavailability of a realistic
simulation framework and high-quality order-level data restricts
researchers' ability to replicate the successes of deep learning in market
making.

Nevertheless, deep learning models (based on CNN architecture)
have done moderately well when it comes to modelling price-formation
mechanisms using order book events as input \citep{Sirignano2019, Cont2019, Tashiro2019, Tsantekidis2017}. The literature is proof that empirical investigation of order-level data has always augmented the model of choice. In short, the ABM, statistical modelling and artificial intelligence approaches to market making all have desirable attributes that the others lack. For example, the CNN architectures lacks self- and cross-excitation effects while modelling LOB dynamics \citep{Zhang2019}. The DLSTM architecture performs hierarchical processing of complex order book events, and as such is able to efficiently capture the temporal structures of LOB \citep{Sagheer2019A, Sagheer2019B}. As market making involves placing optimal bids and asks, the DLSTM architecture, together with the Hawkes process, would efficiently model order arrival
and price-formation mechanisms by constructing a self-exciting multivariate Hawkes process with limit order state process, which are
coupled and interact with each other.


%
\section{Deep Hawkes Process}
\label{sec:DHP_DHP}
In this section, we propose that the Deep Hawkes Process (DHP) be used
to concurrently model the order book event timings and associated event
types. By assimilating it into the order arrival process, the market
makers have control over the sending of different orders at specific
points in time. The basic idea behind our approach is to view the
conditional intensity of the Hawkes process as a nonlinear deterministic
function of past history, and to use DLSTM to automatically learn a
high-dimensional representation from the data. We believe that
capturing the constraints of the Hawkes process and recurrent network
architecture enables a replication of the success of the natural language
process \citep{Du2016, Mei2017} and time series prediction \citep{Sagheer2019A, Sagheer2019B} in market making. 

\subsection{Deep Long Short-Term Memory}

Despite the success of recurrent marked temporal point processes and the
Neural Hawkes process across disciplines, the conventional shallow
recurrent network architecture and its variant long short-term memory
(LSTM) might be inadequate when it comes to modelling noisy
asynchronous order book events. Lately, the DLSTM architecture \citep{Sagheer2019A, Sagheer2019B, Zhu2016} used in action modelling
and multivariate time series forecasting validated its precedence over
traditional LSTM architecture. The architecture of the DLSTM is same
as the previously introduced LSTM, apart from the fact that it involves
multiple LSTM layers stacked on top of each other. 

\subsection{DLSTM-SDAE}

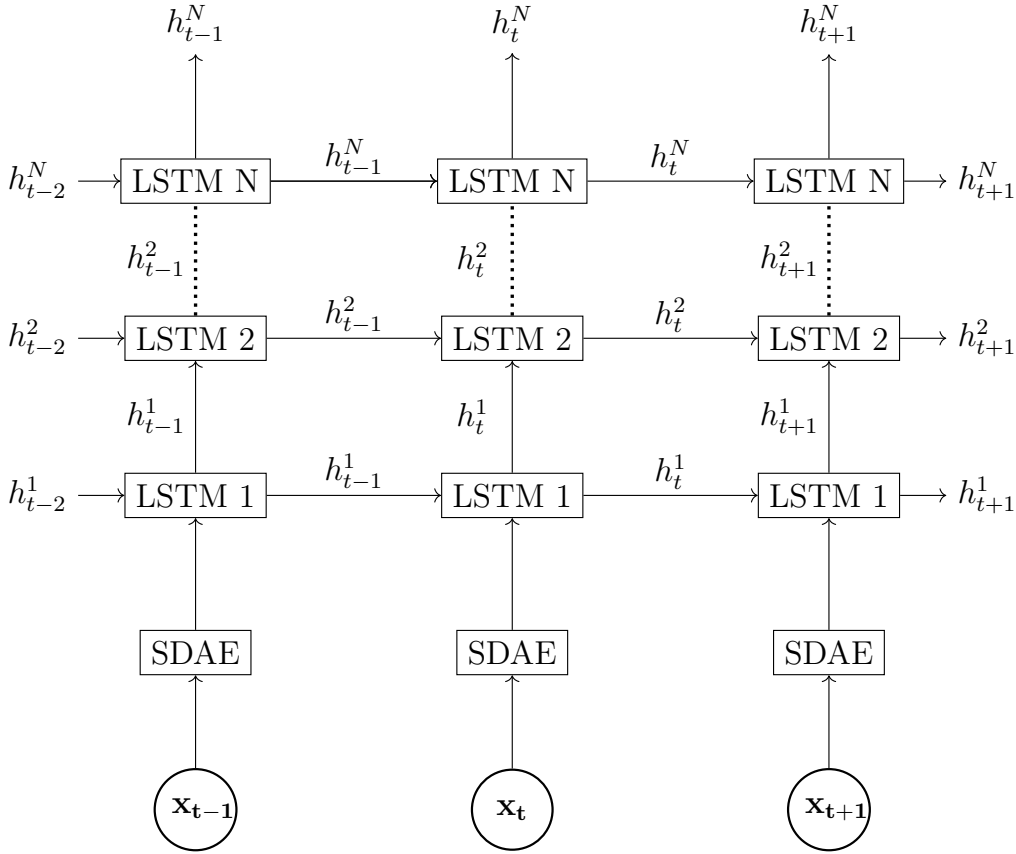
\begin{figure}[H]
\centering
\resizebox{\textwidth}{!}{\begin{tikzpicture}[roundnode/.style={circle,draw,thick, align=center, text width=6mm, minimum size=6mm}]

	\node(h3) at (8,8) {$h^{N}_{t+1}$};
    \node(h2) at (4,8) {$h^{N}_{t}$};
	\node(h1) at (0,8) {$h^{N}_{t-1}$};	
	
	\node[draw] (LSTM9) at (8,6) {LSTM  N};
	\node[draw] (LSTM8) at (8,4) {LSTM  2};
	\node[draw] (LSTM7) at (8,2) {LSTM  1};
    \node[draw] (LSTM6) at (4,6) {LSTM  N};	
	\node[draw] (LSTM5) at (0,6) {LSTM  N};
	\node[draw] (LSTM4) at (4,4) {LSTM  2};
	\node[draw] (LSTM3) at (0,4) {LSTM  2};
	\node[draw] (LSTM2) at (4,2) {LSTM  1};
	\node[draw] (LSTM1) at (0,2) {LSTM  1};
	
	\node[draw] (SDAE1) at (0,0) {SDAE};
	\node[draw] (SDAE2) at (4,0) {SDAE};
	\node[draw] (SDAE3) at (8,0) {SDAE};
	
	\node(H19) at (7.5,5) {$h^{2}_{t+1}$};
	\node(H18) at (7.5,3) {$h^{1}_{t+1}$};
	
	\node(H17) at (10,6) {$h^{N}_{t+1}$};
	\node(H16) at (10,4) {$h^{2}_{t+1}$};
	\node(H15) at (10,2) {$h^{1}_{t+1}$};
	\node(H14) at (2.0,6.3) {$h^{N}_{t-1}$};
	\node(H13) at (2.0,4.3) {$h^{2}_{t-1}$};
	\node(H12) at (2.0,2.3) {$h^{1}_{t-1}$};
	\node(H11) at (3.5,5) {$h^{2}_{t}$};
	\node(H10) at (3.5,3) {$h^{1}_{t}$};
	\node(H8) at (-0.5,5) {$h^{2}_{t-1}$};
	\node(H7) at (-0.5,3) {$h^{1}_{t-1}$};
	\node(H6) at (6,6.3) {$h^{N}_{t}$};
	\node(H5) at (-2,6) {$h^{N}_{t-2}$};
	\node(H4) at (6,4.3) {$h^{2}_{t}$};
	\node(H3) at (-2,4) {$h^{2}_{t-2}$};
	\node(H2) at (6,2.3) {$h^{1}_{t}$};
	\node(H1) at (-2,2) {$h^{1}_{t-2}$};	
	
	\draw[->] (LSTM9) to (H17);
	\draw[->] (H5) to (LSTM5);
	\draw[->] (LSTM8) to (H16);
	\draw[->] (H3) to (LSTM3);
	\draw[->] (LSTM7) to (H15);
	\draw[->] (H1) to (LSTM1);
	\draw[->] (LSTM5) to (LSTM6);
	\draw[->] (LSTM5) to (LSTM6);
	\draw[->] (LSTM3) to (LSTM4);
	\draw[->] (LSTM1) to (LSTM2);
	\draw[->] (LSTM9) to (h3);
	\draw[->] (LSTM6) to (h2);
	\draw[->] (LSTM5) to (h1);
	\draw[dotted, very thick] (LSTM4) to (LSTM6);
	\draw[dotted, very thick] (LSTM3) to (LSTM5);
	\draw[dotted, very thick] (LSTM8) to (LSTM9);
	
	\draw[->] (LSTM7) to (LSTM8);
	\draw[->] (LSTM6) to (LSTM9);
	\draw[->] (LSTM4) to (LSTM8);
	\draw[->] (LSTM2) to (LSTM7);
	\draw[->] (LSTM2) to (LSTM4);
	\draw[->] (LSTM1) to (LSTM3);

	\draw[->] (SDAE3) to (LSTM7);
	\draw[->] (SDAE2) to (LSTM2);
	\draw[->] (SDAE1) to (LSTM1);

	\node[roundnode](x1) at (0,-2) {$\bf{x}_{t-1}$};
	\node[roundnode](x2) at (4,-2) {$\bf{x}_{t}$};
	\node[roundnode](x3) at (8,-2) {$\bf{x}_{t+1}$};
	
	\draw[->] (x1) to (SDAE1);
	\draw[->] (x2) to (SDAE2);
	\draw[->] (x3) to (SDAE3);
	
\end{tikzpicture}
}
\caption{The DLSTM-SDAE architecture.}
\label{fig:DLSTM}
\end{figure}

In an ideal setting, the performance of the DLSTM model proved to be
empirically better than existing contemporary statistical and deep modules. However, as the non-linear variables are modelled in such a way as to be scaled up, the overall learning of the models suffers greatly due to the back-propagation algorithm trapped within multiple local minima \citep{Sagheer2019B}. This may be the biggest hurdle when it comes to modelling limit order book events that comprise complex, asynchronous non-linear multivariate time series. The ultra-noisy order book data also makes processing more challenging. To circumvent the limitations of the conventional stacked LSTM model, we use the stacked denoising auto-encoder (SDAE) together with DLSTM. The SDAE enables the deep neural networks with multiple nonlinear hidden layers to learn complex features from noisy limit order book data \citep{Li2019, Vincent2010} and resolves the random weight initialization of LSTM's units problem in DLSTM \citep{Sagheer2019B}. Figure \ref{fig:DLSTM} shows the proposed DLSTM- based SDAE architecture for DHP.

As shown in Figure \ref{fig:DLSTM}, the reconstructed order book data is  denoised by SDAEs layers in in DLSTM-SDAE architecture. In this method, at the first layer the input $\bf{x_t}$ is corrupted into $\bf{\tilde{x_t}}$ using stochastic mapping $\bf{\tilde{x_t}} \sim \mathcal{S_{\mathcal{D}}}(\bf{\tilde{x_t}}\mid \bf{x_t} )$. Then, the autoencoder maps  corrupted input $\bf{\tilde{x_t}}$ to a hidden representation $\mathit{h} = f_{\theta}(\bf{\tilde{x_t}})$ with encoder $f_{\theta}(\bf{\tilde{x_t}})= (W \bf{\tilde{x_t}} + b)$. Lastly, the decoder $g_{\theta^{'}}$ reconstruct $\bf{z}=g_{\theta^{'}}(\bf{\tilde{x_t}})$ from the hidden representation $\mathit{h}$. The parameters $\theta$ and $\theta^{'}$ are trained using stochastic gradient descent to minimize reconstruction error measured in the squared error loss $L_{2}(\bf{x},\bf{z})=\norm{\bf{x} -\bf{z} }^{2}$. Once mapping is learned, the high-level hidden state $\mathit{h}$ is applied for training the next layer. For detail learning procedure in SDAE, please refer to seminal paper on the subject \citep{Vincent2010}.

At time $t$, the denoised input $\bf{x_t}$ from SDAE is then passed to first layer of LSTM together with previous hidden state $h^{1}_{t-1}$. The hidden state at time $t$, $h^{1}_{t}$ is calculated using recursive LSTM procedure discussed in Equation \ref{eqn:lstm}. Its is then moved to next time step and LSTM layers. In the second layer, the hidden state $h^{1}_{t}$ and the previous $h^{2}_{t-1}$ is used to compute $h^{2}_{t}$ and procedure repeats until last layer is complied.

\subsection{Model Formulation}
\label{ss:MF}

Let $\{t_n, \varkappa_n, k_n\}_{n \in \mathbb{N}}$ be a stream of order book event, where $t_n$ are times of occurrence of an event, its component $\varkappa_n$, and their corresponding mark $\{k_n\}_{n \in \mathbb{N}}$ in $k_n := \{1, \dots , K\}$. Then, the probability that the next event occurs at time $t_n$ is of type $k_n$ is $\mathbb{P}\{(t_n,\varkappa_n, k_n) \mid \mathcal{H}_n, (t_n - t_{n-1})\} dt$. We are interested in the model to predict next event stream $\{t_n, \varkappa_n, k_n\}$ given a past history of event $k_n$, evaluate its likelihood and simulate the next event stream by learning from the past event stream. Equation \ref{eqn:hawkes_mod_a} shows the associated intensity function
of the DHP with relaxed positivity constraints: 

\begin{equation}
\lambda_k(t) = f_{k}(\bf{w}_{k}^{\top}\bf{h}(t))
\label{eqn:hawkes_mod_a}
\end{equation}
The hidden state $\bf{h}(t)$ is updated from the memory sell $\bf{c}(t)$ as in Equation \ref{eqn:hawkes_mod_b}.
 \begin{equation}
  \bf{h}(t) = \bf{o}_{n} \odot \phi (\bf{c}(t)) \text{ for } t \in (t_{n-1}, t_{n}]
  \label{eqn:hawkes_mod_b}
  \end{equation}
  
The life of interval $(t_{n-1},t_n]$ is determined by the next event $k_n$ at $t_n$, DLSTM reads $\{t_n, \varkappa_n, k_n\}$ and updates the current memory cells $\bf{c}(t)$ to $\bf{c}_{n+1}$, associated with hidden state $\bf{h}(t_n)$. The other parameters of the DLSTM are recursively updated according to the Equation \ref{eqn:dlstm}.

\begin{eqnarray}
\label{eqn:dlstm}
\begin{aligned}
\!&{\bf{i}}_{n+1}= \sigma \left( {{\bf{W}}_{xi}}{{\bf{x}}_{n}} + {{\bf{W}}_{hi}}{\bf{h}}(t_n) + {{\bf{W}}_{ci}}{\bf{c}} (t_n) + {\bf{b}}_i \right), \\ 
\!&{\bf{f}}_{n+1} = \sigma \left( {{\bf{W}}_{xf}}{{\bf{x}}_{n}} + {{\bf{W}}_{hf}}{\bf{h}} (t_n) + {{\bf{W}}_{cf}}{\bf{c}} (t_n) + {\bf{b}}_f \right), \\
\!&{\bf{\overline{c}}}_{n+1} = \phi \left( {{\bf{W}}_{xc}}{{\bf{x}}_{n}}\! +\! {{\bf{W}}_{hc}}{\bf{h}}(t_n) \!+\! {\bf{b}}_c \right), \\ 
\!&{\bf{c}}_{n+1} = {\bf{f}}_{n+1}\!\odot {\bf{c}}(t_n)\!+ {\bf{i}}_{n+1}\!\odot {\bf{\overline{c}}}_{n+1}, \\ 
\!&{\bf{\widehat{c}}}_{n+1} = {\bf{\widehat{f}}}_{n+1}\!\odot {\bf{\widehat{c}}} (t_n)\!+ {\bf{\widehat{i}}}_{n+1}\!\odot {\bf{\overline{c}}}_{n+1}, \\ 
\!&{\bf{o}}_{n+1} = \sigma \left( {{\bf{W}}_{xo}}{{\bf{x}}_{n}} + {{\bf{W}}_{ho}}{\bf{h}}(t_n) + {{\bf{W}}_{co}}{\bf{c}}(t_n) + {\bf{b}}_o \right), \\
\!&{\bf{\Bar{\Bar{s}}}}_{n+1} = f \left( {{\bf{W}}_{xd}}{{\bf{x}}_{n}} + {{\bf{W}}_{hd}}{\bf{h}}(t_n) + {{\bf{W}}_{cd}}{\bf{c}} (t_n) + {\bf{b}}_d \right),
\end{aligned}
\end{eqnarray}

where $\bf{x}_n$ is $n^{th}$ input vector represented by one hot encoding of new order book event $k_n$; the activation functions $\sigma \left(x \right)$ / $\phi \left(x \right)$ are sigmoid /  hyperbolic tangent function, respectively; ${\bf{W}}_{A B}$ (e.g., ${\bf{W}}_{c i}$) is the weight matrix from the memory cell to input gate vector; ${\bf{b}}_{B}$ denotes the bias term of $B$ with $B \in \{i,f,c,o,d\}$, $\bf{\Bar{\Bar{s}}}$ is exponential decay parameter and $f(x)=s \log(1+\exp(x/s))$, $s > 0$ is scaled soft plus function. As it can be seen in the Equation \ref{eqn:dlstm}, the parameters are updated using the hidden state $\bf{h}(t_n)$ at time $t_n$, succeeding its decay over interval $t_{n} - t_{n-1}$ rather previous hidden state (Equation \ref{eqn:lstm}). The memory cell ${\bf{c}}(t)$ on the interval $(t_{n-1}, t_{n}]$ follows power-law distribution decaying from ${\bf{c}}_{n+1}$ to ${\bf{\widehat{c}}}_{n+1}$ and defined as:

\begin{equation}
\label{eqn:c_decay}
\bf{c}(t) = {\bf{\widehat{c}}}_{n+1} + \left( {\bf{c}}_{n+1} - {\bf{\widehat{c}}}_{n+1} \right)  \left( \left( t - t_n \right)^{{- \bf{\Bar{\Bar{p}}}}_{n+1}} \right) \text{ for } t \in (t_n,t_{n+1}]
\end{equation}

The DHP, with novel discrete update of stacked LSTM state, allows the model to capture a delayed response, fits non-interacting event pairs, and copes with partially observed event streams. \cite{Mei2017} discuss in detail these benefits of the neural version of the model. In order to ensure mathematical tractability, we have illustrated parameter updates for one of the layers of stacked LSTM, but this can be easily extended to deep architecture. For example, the hidden state ${\bf{h}}^b$ at block $b$ in stacked LSTM is recursively computed from $b = 1\,\text{:}\,N$ and $t = 1\,\text{:}\,T$ using ${\bf{h}}^{b}_{t}= \sigma ({{\bf{W}}_{h^{b-1}h^{b}}}{\bf{h}}^{b-1}_t + {{\bf{W}}_{h^{b}h^{b}}}{\bf{h}}^{b}_{t-1} + {\bf{b}}_h)$.

\subsection{Feedback Loop Exploration}

The empirical results \citep{Morariupatrichi2018, Gonzalez2017} indicate the existence of feedback loop between the order flow and the shape of the LOB, together with the self- and cross-excitation effects. In order to efficiently capture this feedback effect in high-dimensional parameter space, we infuse the feedback loop exploration process into the DLSTM-SDAE architecture, as discussed in Figure \ref{fig:DLSTM}. In connection with designing the deep network architecture and appropriate regularisation, we take into consideration that the network can automatically explore distinct feedback loops for different types of events and their codependency. For example, the feedback effect of market buy and sell orders on price, volume and the bid-ask spread of LOB. 

Consequently, we design a fully connected deep network in which
each neuron represents an LOB state or feedback effect of the preceding
layer, in order to automatically explore the feedback loop. Furthermore,
the neurons in the same layer are partitioned into $\mathfrak{B}$ blocks to take into account different combinations of feedback loops. The corresponding regularisation is incorporated into the loss function and described by:

\begin{equation}
\label{eqn:loss}
\!\mathfrak{L}\!=\!\min_{{\bf{W}}_{xB}}{\mathfrak{L}}\!+{\lambda}_1\!\!\sum_{\substack{B \in S}}{\left\Vert {\bf{W}}_{xB}\right\Vert}_{1}\!\!+\!{\lambda}_2\!\sum_{\substack{B \in S}}\sum_{{\mathfrak{b}}=1}^{\mathfrak{B}}  \left\Vert {{\bf{W}}_{xB,{\mathfrak{b}}}}^T \right\Vert _{2,1}\!,
\end{equation}

where $\mathfrak{L}$ is the loss function of the DLSTM, and other two terms are feedback loop regularization applied to each block in the network \citep{Zhu2016}. ${\bf{W}}_{xB}\!\in {\mathbb{R}}^{N_{N} \times K_{J}} $ is weight matrix, with number of neurons $N_{N}$ and inputs dimension $K_{J}$. The $S$ characterizes the set of gates and cell in LSTM neurons for each block in DLSTM. Lastly, the $\left\Vert {\bf{W}}\right\Vert_{2,1}\!\!=\!\!\sum_i\!\!\sqrt{\sum_j\!\!w_{i,j}^2}$ is a structural $\ell_{21}$ norm. The loss function (Equation \ref{eqn:loss}) was solved by using Adaptive Moment Estimation (Adam) \citep{Kingma2015}. Adam optimization is an augmentation to stochastic gradient descent that is memory efficient, extremely insensitive to hyperparameters, works with sparse gradients, is appropriate for non-stationary objectives, and learns the learning rates itself on a per-parameter basis. It is well suited for highly noisy and/or sparse-gradient order book data.

\subsection{Parameters Estimation}

Given a collection of sequence of order book events $\mathcal{S}$, $\{t_n, \varkappa_n, k_n\}_{n \in \mathbb{N}}$, the the log-likelihood of the model can be expressed as the sum of the log-intensities of lapsed event minus an integral of the aggregate intensities over the whole interval observed till $T$:
\begin{equation}\label{eqn:loglike}
\mathcal{L}\!=\!\  \sum_{n: t_n \leq T}\!\!\log \lambda_{k_n}(t_n)\! - \int_{t=0}^{T}\!\!\lambda(t) dt,
\end{equation} 
    
The parameter $(k,t)$ is estimated maximizing $\mathcal{L}$ using Adam \citep{Kingma2015} and Monte Carlo methods \citep{Mei2017}. The frequently used thinning algorithm adapted from multivariate Hawkes process is then used to sample random sequence from the model.


%
\section{Multi-Agent Simulation Framework}
\label{sec:DHP_MASF}
Multi-agent-based modelling is a bottom-up computational modelling
approach intended to artificially simulate the aggregate behaviours of
systems caused by the actions and interactions between heterogeneous
autonomous agents in an environment or environments \citep{Samanidou2007}. In this section, we describe the important components of multi-agent-based modelling, including environment (market
simulator), agent ecology (trading strategies) and reward (profit and
loss), to study the behaviour of market making agents whose strategies
employ Deep Hawkes processes.

\subsection{Market Simulator}

The market simulator is an essential tool for designing, evaluating and backtesting algorithmic trading strategies under various market scenarios. The market is flooded with various proprietary and open-
source financial simulation frameworks, but the constraints associated
with licences, application interfaces and software design limit their
usability \citep{Maeda2019, Izumi2009}. In this paper, we have designed a multi-asset market simulator from scratch, which is scalable to markets of substantial size. The asynchronous event-based interface is built over realistic market architecture, interface kernels, a matching engine and the Financial Information eXchange (FIX) protocol – an open electronic communications protocol standard used to carry out trades in electronic exchanges. 

\subsubsection{Market Architecture}

The market architecture consolidates the communication interface,
market and matching engine. Figure \ref{fig:DHP_MS} outlines key components of market architecture and their interaction from a high-level perspective. The agents connect to the market via a kernel that hosts order management details. This acts as a transmission channel between agents and markets, thereby providing the extreme throughput and lowest
latency for order transactions. It also throttles the amount of
transactions, as per the market requirement. As such, there is a guarantee of fairness between agents waiting to place orders. The markets
represent an information interchange, in which heterogeneous agents
communicate through kernels for order transactions, processing and
execution according to the matching engine, as per the financial
instruments. The markets respond to order status by sending an
execution report that covers the period from start to market reset event.
This provides opportunities for agents to tweak the parameters in their
trading strategies after every trading period if required.

\tikzstyle{AM} = [rectangle, rounded corners, minimum width=2cm, minimum height=1cm,text centered, draw=black]

\tikzstyle{arrow} = [thick,->,>=stealth]

\begin{figure}[H]
\centering  
\resizebox{\textwidth}{!}{\begin{tikzpicture}[node distance=2.0cm]
\node (A) [AM] {Agents};
\node (K) [AM, right of=A, xshift=1.0cm]{Kernel};
\draw [arrow] (A) -- node[anchor=south] {} (K);
\draw [arrow] (K) -- node[anchor=north] {} (A);
\node (M) [AM, right of=K, xshift=1.0cm]{Markets};
\draw [arrow] (K) -- node[anchor=south] {} (M);
\draw [arrow] (M) -- node[anchor=north] {} (K);
\node (E) [AM, right of=M, xshift=1.0cm]{Matching Engines};
\draw [arrow] (M) -- node[anchor=south] {} (E);
\draw [arrow] (E) -- node[anchor=north] {} (M);
\end{tikzpicture}
}
\caption{Market Simulator.}

\label{fig:DHP_MS}
\end{figure}
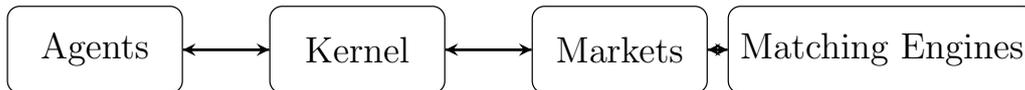

\subsubsection{Communication Interface}
The communication interface between agents and markets follows the
FIX standard protocol to increase efficiency, competition and
innovation. It is an electronic communications protocol designed for the
real-time exchange of information between agents and exchanges,
including agent identifiers, order identifiers, order handling, trade
notifications, broadcasts and execution reports. The widespread
adoption of the FIX protocol across financial markets reduces costs
associated with connectivity, regulatory compliance, liquidity searches
and transactions. Using the FIX protocol, multiple agents can interact
with the market via kernels, simultaneously and independently.

\subsubsection{Matching Engine}
At the core of the market simulator are several \emph{matching engines} for different financial instruments. Each \emph{matching engines} matches bids and asks to execute trades in specific instruments. The orders are matched using \emph{price-time} priority mechanisms. In this context, among bids or asks, the matching algorithms give priority to orders with the highest or lowest price. The ties are broken by giving preference to orders with the earliest submission time compared to other orders. Other well-known priority mechanism algorithms are \emph{pro-rata} and \emph{price-size}  \citep{Gould2013}.

\subsection{Market Ecology}
The success of a financial simulation framework depends on the precise representation of market ecology that can adequately mimic the real
market design. In finance literature, the term "market ecology" \citep{Farmer2002} refers to the composition of heterogeneous trading strategies that keep evolving over time in response to pressure from a contrasting market. The correct mapping of financial market ecology requires the availability of agent-resolved order-level data, which enables identification of sources and events in the market. In the absence of agent-resolved data, the trading strategies are classified by theoretical considerations, the results of surveys, direct investigations of the trading profile of classes of investors and proxies for HFT \citep{Kirilenko2017, Mankad2013}. In this paper, we adapt the market ecology from the different strands of academic literature \citep{McGroarty2019, Paulin2019, Musciotto2018, Kirilenko2017, Leal2016, Mankad2013, Toth2012}. 
 
\subsubsection{Deep Hawkes Market Makers}

Despite the existence of sophisticated market making strategies, the classical "buy low and sell high" strategy \citep{Cartea2014} is a preferred strategy to make money in the securities market. The success of making a profit from short-term price predictions on bids and asks hinges on placing orders at precisely the right time. The mathematical modelling of the complicated order arrival process is done using the Poisson process \citep{Chakraborti2011E}, which assumes that orders arrive
randomly. However, this is contrary to empirical findings, which show that order arrival times are strongly connected \citep{Rambaldi2017}, have self- and cross-excitation effects \citep{Morariupatrichi2018} and that a feedback loop exists between the order arrival and the state of the LOB \citep{Gonzalez2017}. In this article, we incorporate the feedback loop between order arrival and the state of the limit order book, together with self- and cross- excitation effects, using DHP, to design a high-frequency market making strategy.

The order book events in the securities market are stochastically excited or impeded by a pattern in the past event streams. The market makers are interested in learning the distribution and structure of order book events stream to accurately predict the next order (limit orders, market orders, cancellations, etc.) together with an associated labels (price, volume, etc.). Given a stream of order book events $\{t_n, \varkappa_n, k_n\}_{n \in \mathbb{N}}$, the market makers calculate the probability that the next event occurs at time $t_n$ is of type $k_n$ and its probability density  conditioned on the history of events $\mathcal{H}_n$ by:
\begin{equation}
\lambda(t)dt = \mathbb{P}\{(t_n,\varkappa_n, k_n) \mid \mathcal{H}_n, (t_n - t_{n-1})\}.
\end{equation}

\begin{equation}
f_n(t) = \lambda(t) \exp \left( -\int_{t_{n-1}}^{t} \lambda(\tau) d\tau \right).
\end{equation}

To predict the time and the next event having minimum loss without information about the time $t_n$, we choose $ \hat{t}_{n} = \int_{t_{n-1}}^{\infty} t f_n(t) dt$ and $\hat{k}_{n} = \argmax_{k} \int_{t_{n-1}}^{\infty} \frac{\lambda_k(t)}{\lambda(t)} f_n(t) dt$. The associated intensity function for calculating the next order book events is the same as DHP as described in Equation \ref{eqn:hawkes_mod_a}. 

The real securities market has numerous distinct order book events, which is difficult to incorporate in our simulation framework. For the
sake of computational tractability, we decided to only include limit order buy/sell, market order buy/sell, and partial/full cancellations. We also assume that the high-frequency market maker is trading a single security in a market whose price at time $t$ is denoted by $p_t$. Unlike traditional market making strategies \citep{Spooner2018}, the DHP market makers can trade with rational limit or market order quantity, price surge and bid-ask spread distributions. 

The deep hawkes market maker (DHMM) place orders at a specified depth relative to the mid-price, $\overline{p_t}$. At each time step $t$, the DHMM agent's pricing mechanism is given by:

\begin{equation}
p^{a,b}_t = \overline{p_t} +  \sum_{i=1}^{\overline{\delta}}i\cdot\mathbf{J}^{i,u}_t - \sum_{i=1}^{\overline{\delta}}i\cdot\mathbf{J}^{i,d}_t 
\end{equation}

where $\overline{p_t}$ is the mid price at time $t$, $\mathbf{J}^{i,u}$ is the number of upward jumps with $i$ ticks, and $\mathbf{J}^{i,d}$ is the number of downward jumps with $i$ ticks between $0$ and $t$, $i = 1,\dotsc,\overline{\delta}$. The intensities of $\mathbf{J}^{i,u}$ and $\mathbf{J}^{i,d}$ are $\lambda_{k,u}(t)$ and $\lambda_{k,d}(t)$, respectively,

\begin{equation}
\lambda_{k,i}(t) = f_{k,i}(\bf{w}_{k,i}^{\top}\bf{h}(t)), \; \;  k=u,d.
\label{eqn:hawkes_mod_A}
\end{equation}

The parameters of the Equation \ref{eqn:hawkes_mod_A} are calculated as discussed under model formulation in Section \ref{ss:MF}.

Most of the quantitative finance research into the high-frequency market making problem is based on the assumption of constant order size \citep{Huang2015}. However, the empirical analyses suggests that the order sizes have striking statistical distribution at different timescales \citep{Lu2018, Rambaldi2017, Mu2009}. The limit order size follows $\mathfrak{q}$-Gamma distribution \citep{Mu2009}. The market maker's willingness to sell or buy specified quantities of securities is defined as:    

\begin{equation}
 q^{a,b}_{lt}= \left[\Gamma (\alpha, \beta)\right]^{q_{max}}_{q_{min}} \cdot \left(\frac{I_t \pm \bar{I}}{\bar{I}}\right),
 \label{Eq:G}
\end{equation}
where $I_t$ is inventory at time $t$, $\bar{I}$ maximum inventory, and $\Gamma (\alpha, \beta)$ is $\mathfrak{q}$-Gamma distribution is described as:

\begin{equation*}
 \Gamma (\alpha, \beta; q)= \frac{1}{Z} \left(\frac{q}{\alpha}\right)^{\beta}
 \left[1-(1-\mathfrak{q}){\frac{q}{\alpha}}\right]^{\frac{1}{1-\mathfrak{q}}}~,
 \label{Eq:qGamma}
\end{equation*}

\begin{equation*}
 Z=\int_{0}^{\infty}\left(\frac{q}{\alpha}\right)^{\beta}
 \left[1-(1-\mathfrak{q}){\frac{q}{\alpha}}\right]^{\frac{1}{1-\mathfrak{q}}}{\rm{d}}q.
 \label{Eq:Z}
\end{equation*}

One striking feature of equity markets is the existence of short- lived
limit orders that are modified or cancelled once every 50 milliseconds \citep{Dahlstrm2018}. The limit order cancellation is an important
characteristic of market making strategies that are related to expected
profit, bid-ask spread and order queue position. We model cancellation
sizes as follows:

\begin{equation}
 q^{a,b}_{ct}= \left[P_c(q;Q)\right]^{q_{max}}_{q_{min}} \cdot \left(\frac{I_t \pm \bar{I}}{\bar{I}}\right),
 \label{Eq:TG}
\end{equation}
where $P_c(q;Q)$ is truncated geometric distribution \citep{Lu2018}. The LOB is represented as $[Q_{-i}:i = 1,\ldots,L]$ and $[Q_{i}:i=1,\ldots,L]$ with corresponding quantities $q_i$. The truncated geometric distribution is defined as:
\begin{equation*}
P_c(q;Q)= \mathbb{P}[q|Q] = \frac{p_{c}^{0}(1-p^{0}_{c})^{q-1}}{1-(1-p^{0}_{c})^Q}\mathbbm{1}_{\{q\leq Q\}}.
\end{equation*}

Finally, the market order follows a mixture of truncated geometric distribution and the dirac delta distribution \citep{Lu2018}. The market order size that a market maker is willing to buy or sell is described as: 

 \begin{equation}
 q^{a,b}_{mt}= \left[P_m(q;Q)\right]^{q_{max}}_{q_{min}} \cdot \left(\frac{I_t \pm \bar{I}}{\bar{I}}\right),
 \label{Eq:TGD}
 \end{equation}
  
 \begin{align*}
 P_m(q;Q) &= \theta_0\frac{p_m^0(1-p_m^0)^{q-1}}{1-(1-p_m^0)^Q}\mathbbm{1}_{\{ q\leq Q\}} \\
 &+ \sum_{k=1}^{\lfloor \frac{Q-1}{5} \rfloor} \theta_k\mathbbm{1}_{\{q=5k+1\}}+\theta_\infty \mathbbm{1}_{\{q=Q, Q \neq 5n+1\}}.
 \end{align*}

The parameters $\{p_c^0, p_m^0, \theta_0, \theta_k, \theta_\infty\}$ are estimated using a maximum likelihood method. The details of estimation and calibration can be retrieved from the \cite{Lu2018}. The market orders are used to clear the unexecuted inventory at the end of trading.

\subsubsection{Probabilistic Market Makers}
To ensure fair competition with DHMM and incorporate the existing state-of-the-art, we include a probabilistic estimate-based benchmark strategy \citep{Das2005} adapted to our simulation framework. In this market making strategy, the agent attempts to track the fundamental price of securities by maintaining a probability density estimate of the fundamental price.

The probabilistic market makers (PMM) intent to sell or buy $q$ unit of security at time $t$ for price $p^{a,b}_t$ in a market populated with uninformed, informed and noisy informed agents. Let us assume that the fundamental price of the security at time $t$ is $f_t$, $\xi$ be the fraction of informed agents and the probability of buy or sell orders by the uninformed agents is $\zeta$. The noisy informed agents assumes that the price of securities follow normal distribution $p_t \; = \; f_t + \mathcal{N}_s(0,\sigma^{2}_{n})$. Whereas the fundamental price of security evolves according to a jump process. The order book event defines the jump and prices follow normal distribution. The PMM ask and bid prices at time $t$ are then defined as:

\begin{equation} 
\begin{split}
p^{a,b}_{t} & = \frac{1}{P_{Buy, Sell}}\sum_{f_{t}=f_{min}}^{f_{t}=p^{a,b}_{t}} \left[ \left( (1 - \xi) \zeta + \mathsf{Pr}(\mathcal{N}_s(0,\sigma^{2}_{n})\gtrless(p^{a,b}_{t} - f_{t})) \right) f_{t} \mathsf{Pr}(f = f_t) \right]  \\
 & + \frac{1}{P_{Buy, Sell}}\sum_{f_{t}=p_{t}^{a,b}+1}^{f_{t}=f_{max}} \left[ \left( (1 - \xi) \zeta + \mathsf{Pr}(\mathcal{N}_s(0,\sigma^{2}_{n})\lessgtr(f_{t} - p^{a,b}_{t})) \right) f_{t} \mathsf{Pr}(f = f_t) \right] 
\end{split}
\label{Eq:PMM_ab}
\end{equation}

\begin{equation*} 
\begin{split}
P_{Buy, Sell} & = \sum_{f_{t}=f_{min}}^{f_{t}=p^{a,b}_{t}} \left[ (1 - \xi) \zeta + \mathsf{Pr}(\mathcal{N}_s(0,\sigma^{2}_{n})\gtrless(p^{a,b}_{t} - f_{t})) \right] \mathsf{Pr}(f = f_t)   \\
 & + \sum_{f_{t}=p_{t}^{a,b}+1}^{f_{t}=f_{max}} \left[ (1 - \xi) \zeta + \mathsf{Pr}(\mathcal{N}_s(0,\sigma^{2}_{n})\lessgtr(f_{t} - p^{a,b}_{t})) \right] \mathsf{Pr}(f = f_t) 
\end{split}
\label{Eq:P_ab}
\end{equation*}

where $P_{Buy, Sell}$ is a priori probability of a buy or sell order and $\mathcal{N}_s(0,\sigma_{n})$ is sample from normal distribution. The bids/asks equations derivation, its approximate solutions, density estimate update, and algorithm are discussed in the benchmark paper \citep{Das2005}. We add layers of complexity on the benchmark algorithm by allowing the PMM to sample order or cancellations size from a normal distribution. The order cancellations size at time $t$ determined as follows:

\begin{equation}
q_{t;o,c}^{a,b} = \eta_{o,c}(\bar{p}_{t}-\bar{p}_{t-1})+\mathcal{N}_s(0,\sigma^{2}_{o,c}), \; \; 0 < \eta_{o,c} < 1
\label{Eq:P_oc}
\end{equation}
\subsubsection{Fundamental Traders}

Fundamental traders decide to trade based on the presumption that the
securities prices will eventually return to their basic, intrinsic or
fundamental value. Therefore, they strive to buy (sell) the security when
the price at time t is below (above) its fundamental value. Fundamental
traders are predominantly categorised as buyers or sellers, depending on
the inventory at the end of a trading day. The accumulation of directional net positions is an important element in identifying buyers or sellers, since the latter acquire sizable net positions by executing numerous small-size orders, while the former only execute a couple of large orders \citep{Kirilenko2017, Mankad2013}. According to the agent ecology literature, the fundamental traders assume that fundamental value of a security will follow a random walk:

\begin{equation}
f_t = f_{t-1}(1+\delta_f)(1+x_t), \; \; \; \delta_f > 0; \;  x_t  \sim \mathcal{N}(0, \sigma^{2}_x)
\end{equation}

Given last mid-price at time $t$, the limit order price by fundamental traders is determined by:

\begin{equation}
p_t = \bar{p}_{t-1}(1+\delta_f)(1+z_t), \; \; z_t  \sim \mathcal{N}(0, \sigma^{2}_z)
\end{equation}

Finally, the order under fundamental traders strategy are calculated as follows:

\begin{equation}
q_{t;f} = \eta_{f}(f_{t}-\bar{p}_{t-1})+\mathcal{N}_s(0,\sigma^{2}_{f}), \; \; 0 < \eta_{f} < 1
\end{equation}

The decision to buy or sell is governed by following logic:

\begin{equation}
    \mathcal{D}_t= 
\begin{cases}
    Buy ,& q_{t;f} \geq 0\\
    Sell,              & q_{t;f} < 0
\end{cases}
\end{equation}

\subsubsection{Chartist Traders}
Unlike fundamental traders, the chartist or technical trader's strategy depends on predicting future price direction based on past price movement. The chartist traders in our simulation framework use a simple trend-following strategy described in \cite{Leal2016}. The price, order size and trade direction are described below:

\begin{equation}
p_t = \bar{p}_{t-1}(1+\delta_c)(1+z_t), \; \; z_t  \sim \mathcal{N}(0, \sigma^{2}_c)
\end{equation}

\begin{equation}
q_{t;c} = \eta_{c}(\bar{p}_{t-1}-\bar{p}_{t-2})+\mathcal{N}_s(0,\sigma^{2}_{c}), \; \; 0 < \eta_{c} < 1
\end{equation}

\begin{equation}
    \mathcal{D}_t= 
\begin{cases}
    Buy ,& q_{t;c} \geq 0\\
    Sell,              & q_{t;c} < 0
\end{cases}
\end{equation}

\subsubsection{Noise Traders}
In the securities market, noise traders make trading decisions based
solely on non-information. In the models, they serve as an essential
proxy for randomness, no trade and no speculation. We incorporate the
slightly more evolved noise or background traders from a seminal paper by \cite{Wah2017}. The noise traders ask or bid price is determined by its fundamental private valuation and trading strategy. The fundamental value evolves according to a mean-reverting stochastic process \citep{Wah2017}.

\begin{equation}
f_t = max\left[0, \eta_{n}\bar{f}+\eta_{n}(1-f_{t-1})+y_t \right], \; 0 < \eta_{n} < 1; \; y_t  \sim \mathcal{N}(0, \sigma^{2}_n) 
\end{equation}

The private valuation for the noise traders at time $t$ is given by:
\begin{equation}
p_v = max \left[ 0, d_f\right], \; d_f \sim \mathcal{N}(f_t, \sigma^{2}_v)
\end{equation}

The noise trader calculates its private value and decide to buy or sell $q_{t;n}$ order sampled from a normal distribution,$\mathcal{N}(0, \sigma^{2}_n)$, with equal probability of 1/2. 

\subsection{Reward Design}
Unlike traditional reward design, in which an agent's performance is assessed at the end of a trading period, we calculate the agent's
instantaneous rewards at each timestep $t$. The reward function for the agents ($j$) comprises profit \& loss (PnL), inventory cost (IC) and transaction cost (TC). The PnL is simple profit or loss made by the agents through buying or selling security at the exchange. Its is defined as:

\begin{equation}
PnL_{t;j} = q^{a}_{t;j}\left(p^{a}_{t;j} - \bar{p}_{t} \right)+q^{b}_{t;j}\left(\bar{p}_{t} - p^{b}_{t;j} \right)
\end{equation} 

As a agent's inventory is exposed to the volatility of the market price, we incorporate it our reward design using a term associated with inventory cost. Its given by:

\begin{equation}
IC_{t;j} = I_{t;j}\left(\bar{p}_{t} - \bar{p}_{t-1} \right)
\end{equation} 

Finally, we consolidate a quadratic penalty on the number of shares executed to account for transaction cost. Specifically, the transaction cost for order executed $q_{t}^{e}$ by agent $j$ till time $t$ is:

\begin{equation}
TC_{t;j} = \daleth \left( q_{t;j}^{e}\right)^{2}, \; 0<\daleth<1
\end{equation}

The reward function is the sum of orders bought or sold plus inventory cost less a transaction cost penalty.

\begin{equation}
R_{t;j} = PnL_{t;j}+IC_{t;j}-TC_{t;j}.
\end{equation}

\subsection{Capital Allocation}
\label{sec:CA}

The amount of currency units held by an agent is represented by capital. Prior to securities market opening in simulation framework, every heterogeneous agents endowed with different amount of capital by a power law distribution. The agent's initial capital $c_a$ follows a power law if it is drawn from drawn from a probability
distribution $p(c_a) \propto {c_a}^{- \alpha_a}$. The $\alpha_a$ is referred as scaling parameter which ordinarily lies between 2 and 3 \citep{Clauset2009}.


%
\section{Experiments}
\label{sec:DHP_E}
In this section, we elaborate on data, its processing, performance
metrics, benchmarks, training and the parameter configuration for the
proposed model.
\subsection{Data}

We use the publicly available historical Nasdaq TotalView-ITCH 5.0 data feed sample \footnote{\url{ftp://emi.nasdaq.com/ITCH/Nasdaq_ITCH/}} to reconstruct limit order book \citep{Huang2011}. The reconstructed database provides tick-by-tick details of full order book depth by listing every quote and order at each price level of a specific security in Nasdaq, NYSE, and regional-listed securities on Nasdaq. The raw data feed in the binary format has a series of sequenced messages to describe the system, securities, order, and trade events at a resolution of the nanosecond scale. The event stream at nanosecond timestamp guarantees the inclusion of stochastically missing events which might increase the predictive accuracy of the Deep Hawkes model. Although the neural hawkes model \citep{Mei2017} is expressive enough to take account of missing event stream, it makes sense to access the performance of deep hawkes model at millisecond resolution order book data as compared to nanoseconds. Nasdaq uses multiple messages to indicate the current order, trading, system, and circuit breakers event's status as discussed in technical report \citep{Nasdaq2020}. For mathematical tractability, we have sampled high frequency data for hundred most liquid securities over eight days from reconstructed orderbook. The extracted sample data consists of approximately a billion transaction records at nanosecond resolutions together with the possible event of limit order buy/sell, market order buy/sell, and cancellations partial/full. The reconstructed limit order book for Apple at 11:21 on March 29th, 2018 is shown in Figure \ref{fig:DHP_LOB}.  

The reconstructed orderbook data is divided into training, validation and test set. For a single security at nanosecond and millisecond resolution, the descriptive statistics are given in Table \ref{tab:DHP_DS}. The validation set is included to optimize the model's hyper-parameters while training, thus having control at over-fitting. To avoid high variance in the data set, we only record the average value over multiple splits denoted by $\approx$ in the Table \ref{tab:DHP_DS}.
 
\begin{table}[H]
\caption{Descriptive statistics of the orderbook data}\label{tab:DHP_DS}
 \resizebox{\textwidth}{!}{ \begin{tabular}{clrrrrr}
    \toprule
    \toprule
    \multirow{2}{*}{Data} & 
      \multicolumn{3}{c}{\# Orderbook Event Token} &
      \multicolumn{3}{c}{Stream Length} \\
      \cmidrule(l){2-4} \cmidrule(l){5-7}
    & Train & Val & Test & Min & Mean & Max \\
    \midrule
    Nanosecond & $\approx 9210480$ & 921000 & 2302620 & 26752 & 85874 & 116217  \\
    \midrule
    Millisecond & $\approx 432116$ & 42252 & 108029 & 1167 & 3670 & 5216 \\
    
   \bottomrule
    \bottomrule
  \end{tabular}
  }
  
\end{table}

\subsection{Performance Metrics}

DHMM agents use DHP to accurately predict Experiments order book events (buy, sell or cancel) and their timing. The accuracy of DHMM's predictions in terms of events and time is an important determinant of its trading profitability. The performance metrics are vital components for measuring the performance of the trained model's prediction reliability with observed test data. The widely used scale-
dependent metric\citep{Hyndman2006}, root mean square error (RMSE) and classification error rate (ER) were used to evaluate the prediction performance of the Neural Hawkes model. Following \cite{Mei2017}, we predict each prevailed order book event stream $\{t_n, \varkappa_n, k_n\}$ from the past event stream $\mathcal{H}_n$ and evaluate prediction using RMSE and ER.

The predominant metric for evaluating the performance of the agents
is profit and loss at the end of the trading period. However, this approach may be misleading, as agents are tested across heterogeneous securities, with varying pricing and liquidity structures. Alternatively, to efficiently capture spread, we use a normalised PnL (NPnL) with inventory and quadratic transaction costs. The NPnL is calculated every hour by dividing the total reward by the weighted average market spread. To take account of the small inventories maintained by the market maker,, \cite{Spooner2018} introduced the mean absolute position (MAP)
metric. An extreme score under this metric indicates a risky speculative
strategy, while a moderate one indicates a strategy based on a stagnant
market. We record the variability for NPnL and MAP using the standard
deviation and mean, respectively.

\subsection{Benchmarks}
We aim to evaluate the performance of the DHMM with a modified
probabilistic estimate-based benchmark strategy \citep{Das2005}. The
market making strategy is an extension of the classic information-based
model \citep{Glosten1985}, in which agents use the probability
estimates of the fundamental price of securities to set bid and ask prices. The agents can sample limit, market or cancellation orders from the normal distribution or contradictory to a unit market order. We
implement the probabilistic estimate-based strategy at the top of our
simulation framework in continuous-time simulation rather than a
discrete-time simulation. This provides the perfect test-bed to assess the performance of the simulation framework in extending the discrete-time mechanisms to continuous-time, where heterogeneous agents interact
asynchronously. 

The DHP extends the seminal Neural Hawkes process to the deep
learning framework in a market making setting. We introduce the novel
architecture to circumvent complications related to random weight
initialisation, training and noisy order-level data \citep{Sagheer2019B}. Given that the Neural Hawkes process is the kernel of our proposed deep model, we evaluate the performance of market making agents that use the earlier model in their trading strategies. For comparison purposes, we use the same architecture and training mechanism as discussed in the seminal paper \citep{Mei2017}. The neurally self-modulating multivariate Hawkes process also acts as benchmark model for evaluating DHP’s performance on the prediction of order book events and in terms of time on the reconstructed limit order book data.

\subsection{Training}

The high-frequency marker making agents uses DHP to learns from reconstructed limit orderbook data to place bids or asks or cancels at suitable time. The learned prediction is then infused into the market making strategy to trade with the simulation framework. The agents learn the system parameters in a two-step process. Firstly, the preprocessed order book stream, n-th event, $k_n$ is embedded into a latent space before passing into SDAE layer together with timing $t_n$. The deep network, consists of a stack of multiple DAEs, generate higher representation of convoluted order book events interaction. The high level denoised representations are then fed into DLSTM to predict the next order's type and the time to evaluate the loss. The DLSTM-SDAE learns the deep representation in two phases: pre-training and fine-tuning. In pre-training, a greedy layer-wise structure is used to train each layer of DAE iteratively, to form a three-layer SDAE. At the end of pre-training, a stack of three LSTMs is produced as an output of SDAE. Secondly, the parameter of DLSTM-SDAE is then fine-tuned to minimise the error in predicting events and time, using using stochastic gradient-descent and Adam optimisation algorithms. The early stopping methods used on the validation set’s log-likelihood performance were also used on the held-out validation set to avoid overfitting. We also add isotropic Gaussian noise to augment generalisation in the performance of the events' classification. Table \ref{tab:DHP_PC} lists the hyper-parameters tuned by validation set performance for the DLSTM-SDAE network architecture. The other non-LSTM parameters includes  $s_n \in \mathbb{R}$ and $\mathbf{W}_n \in \mathbb{R}^D$ as discussed in Section \ref{sec:DHP_DHP}. The market making agent using NHP uses single layer LSTM  and the number of hidden nodes from a small set $\left(64, 128, 256, 512, 1024 \right)$ as described in the base paper \citep{Mei2017}. The hyperparameters are optimized based on the performance of the validation set.

The high-frequency market making agents are trained in the simulation framework for 1000 trading days. Each trading day starts at 9:30 and lasts until 16:00. Two hundred trading days were used to fine tune the hyper-parameters using random search. We acknowledge the existence of
differences between the real data and simulated data, but firmly believe
that they are generated from the same mechanisms – a claim substantiated by agent-based models that reproduce stylised facts similar to empirical findings. Taking the above into consideration, we train market makings agents using DHP and NHP for 100 trading days five times. The aim of this exercise is to synchronise the agents' learning over different sets of data generated from the same stochastic process. We then test the performance of the agents against the benchmark for 300 trading days. To ensure fair competition with market making agents, we use heterogeneous market ecology consisting of fundamental, chartist and random agents. The important parameters pertaining to the trading agents in the simulation framework are given in Table \ref{tab:DHP_PC}.  

\begin{table}[H]
    \centering
    \caption{Parameters configuration.}\label{tab:DHP_PC}
  \resizebox{\textwidth}{!}{  \begin{tabular}{lll}
        \toprule\toprule
        Description  & Parameter/Hyperparameter  & Value \\
        \midrule
        Total sessions size & $\mathsf{T}_{total}$ & 1300 days \\
        Training sample size & $\mathsf{T}_{train}$ &  1000 days \\
        Testing sample size & $\mathsf{T}_{test}$ & 300 days \\
        \midrule
        Total number of traders &$a_t$& $  \approx 10^4$ \\
        Number of market makers &$a_m$& 3 \\
        Number of fundamental traders  &$a_f$& 3000 \\
        Number of chartist traders  &$a_c$& 6000 \\
        Number of noise traders  &$a_n$& 4000 \\
        
        \midrule
        Initial capital & $\mathsf{c}_{a}$ & $ \sim p(2.3) \; \times \; 10^4$ \\
        Min inventory & $\min \Inventory$ & $  \sim - p(2.3) \; \times \; 10^5$  \\
        Max inventory & $\max \Inventory$ & $  \sim p(2.3) \; \times \;10^5$ \\
        \midrule
        Scale factor & $s_n $ & 1 \\
        LSTM weights & $\mathbf{W}_n$ & $\sim \mathcal{N}(0,0.01) $\\
        DHMM limit order parameters & $\Gamma (\alpha, \beta)$ & $\Gamma (0.07, 1.52)$\\
        DHMM limit order cancellations parameters &$P_c(q;Q)$ & 0.60\\

        Fraction of informed agents & $\xi$ & 0.33\\
        Probability of buy/sell by uninformed agents &$\zeta$ & 0.33\\
        PMM order cancellation size parameter & $\eta_{o,c}$ & 0.04\\
        Fundamental order size parameter & $\eta_{f}$ & 0.04\\
        Chartist order size parameter & $\eta_{c}$ & 0.04\\
        Noise price parameter &$\eta_{u}$& 0.04 \\
        Transaction cost penalty & $\daleth$& 0.06\\

        \midrule 
        Number of layers (DAE/LSTM) & $N_{L}$ & 3/3\\
        Number of hidden unit per layer & $N_{H}$1024\\
        Learning rate for pretraining & $\alpha_{LPT}$ &0.05 \\
        Learning rate for fine-tuning &$\alpha_{LFT}$& 0.10\\
        Number of pretraining epochs & $\epsilon_{PT}$&100\\
        Additive isotropic Gaussian noise & $\eta_{noise}$& $\sim \mathcal{N}(0,0.50)$ \\
            
        \bottomrule\bottomrule
        
    \end{tabular}}
\end{table}


%
\section{Results}
\label{sec:DHP_R}
In this section, we investigate the performance of market making agents
in predicting types of order book events and their timestamps. Having
learned which orders to send and at what time, we evaluate the agent’s
trading performance in the simulation framework. We tweak order
cancellations to examine the impact on the agent’s profitability and the
microstructure of the order book. We then check the robustness of the
model by performing sensitivity analysis. Finally, we validate our
simulation framework by reproducing stylised facts with our simulated
data.

\subsection{Predictive Performance}
\begin{figure}[H]
    \centering
    \begin{subfigure}[t]{0.32\textwidth}
        \centering
        \includegraphics[width=\textwidth]{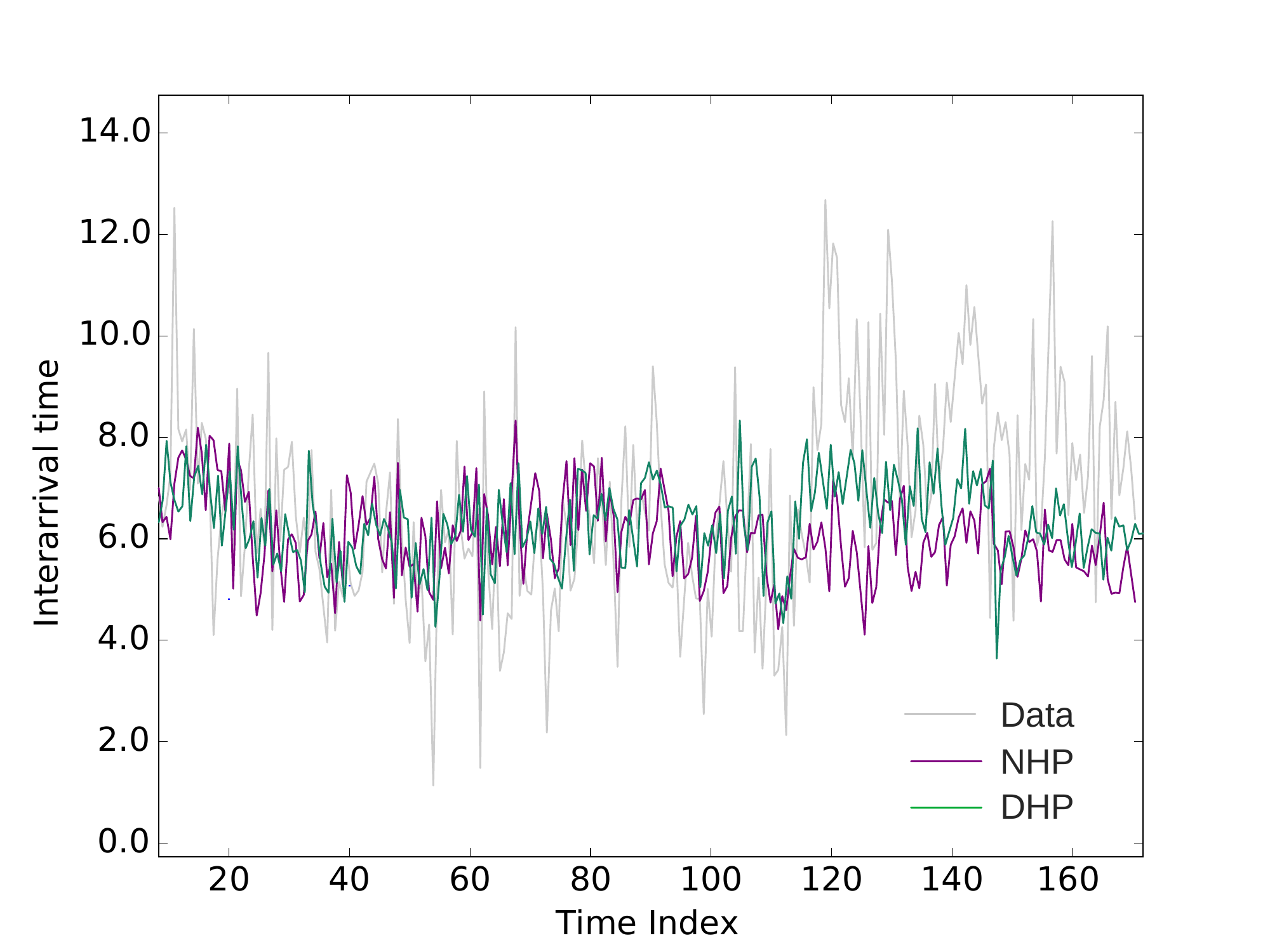}
  \caption{Time Prediction Graph}\label{fig:IETN}
      \end{subfigure}
    \begin{subfigure}[t]{0.32\textwidth}
        \centering
        \includegraphics[width=\textwidth]{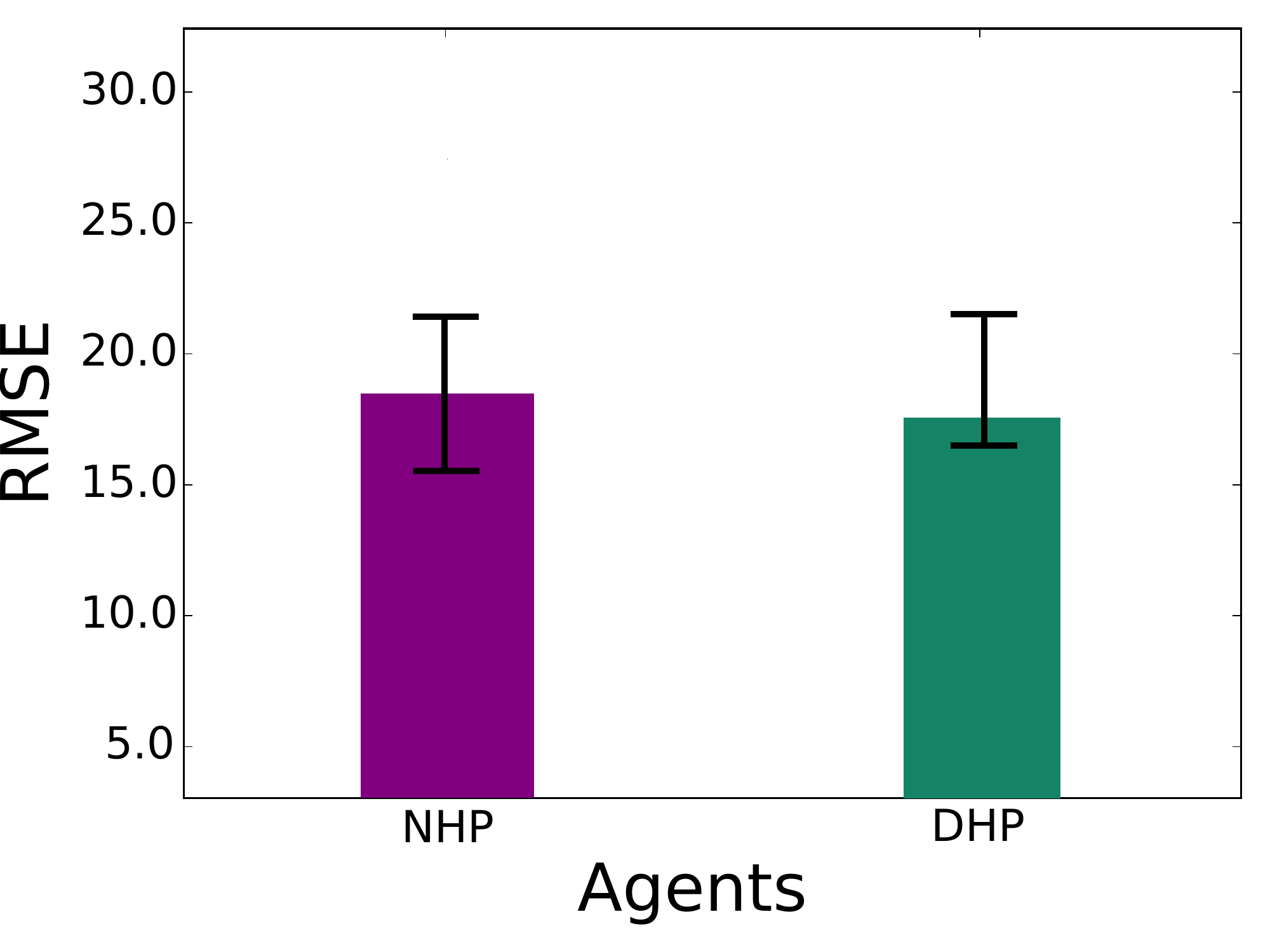}
  \caption{Time Prediction Error}\label{fig:RMSEN}
    \end{subfigure}
    \begin{subfigure}[t]{0.32\textwidth}
        \centering
        \includegraphics[width=\textwidth]{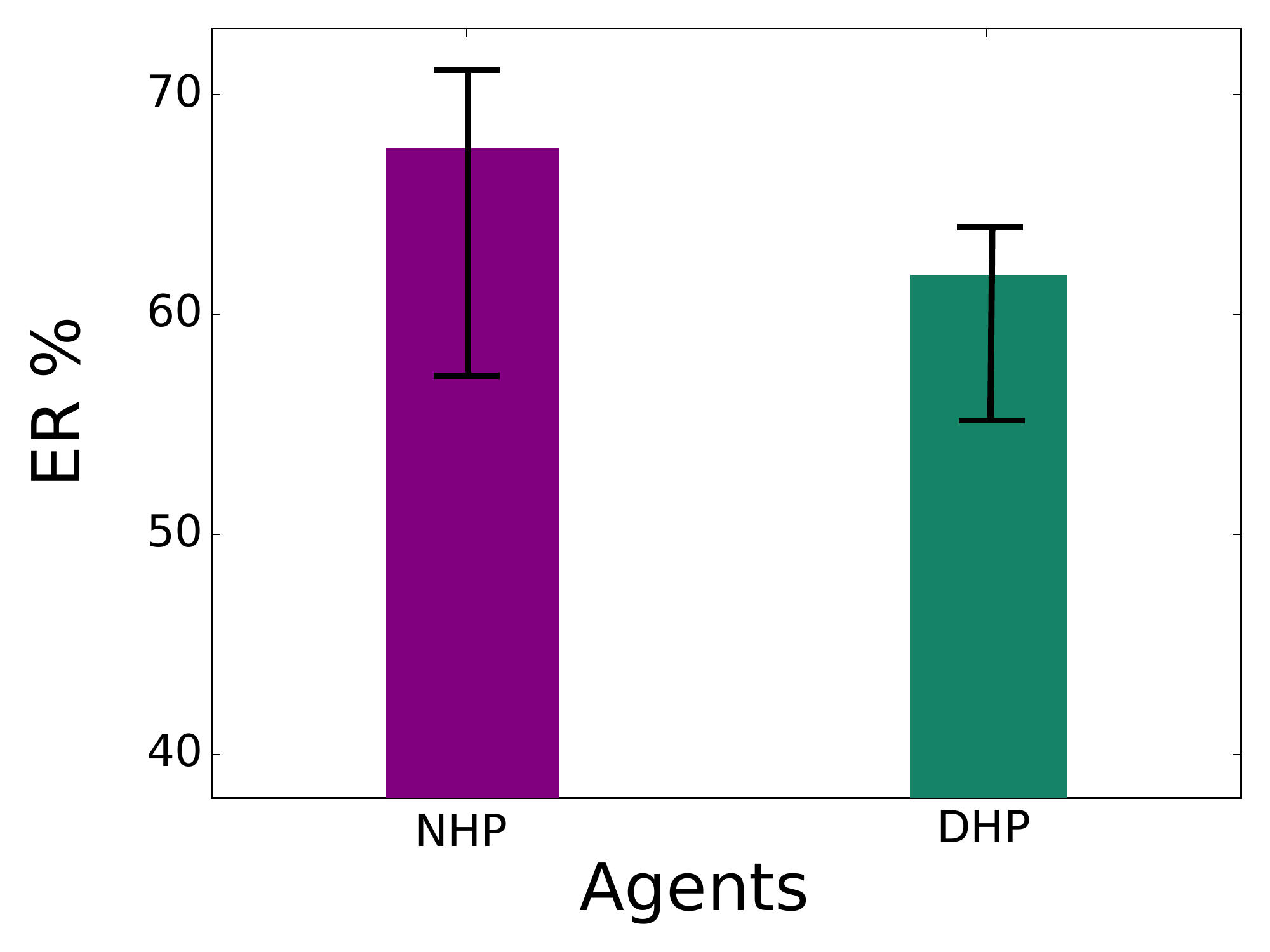}
  \caption{Event Prediction Error}\label{fig:ERN}
      \end{subfigure}
          
   \caption{Performance evaluation of high-frequency market making agents in predicting order book events and time at nanosecond resolution. The standard deviation over 10 experiments using different train-val-test sample is denoted by error bar. }\label{fig:PTEN}
   \end{figure}

Given a stream of order book events $\{t_n, \varkappa_n, k_n\}_{n \in \mathbb{N}}$, the market makers seek to predict the next event type and its time. We evaluate predictive performance of $\hat{t}_{n}$ and $\hat{k}_{n}$ using RMSE and ER, respectively.To avoid getting entangled in the problem of overfitting, we divide the training set into the sub-training and validation sets. We train DHP and NHP models on the sub-training set so as to choose hyperparameters for validation set. Following the training procedure of \cite{Mei2017}, we generate the predictive performance of the market making agents on reconstructed order book data at nanosecond resolution in Figure \ref{fig:PTEN}. As is evident from the figure, neither model is invariably better at predicting events or, in particular, time. It seems that the both models do not explicitly address the complex dynamics of asynchronous order book data at nanosecond resolution. The event dynamics at nanosecond timestamps need much more sophisticated models to filter noise and to model event interaction and non-linearity.

\begin{figure}[H]
    \centering
    \begin{subfigure}[t]{0.32\textwidth}
        \centering
        \includegraphics[width=\textwidth]{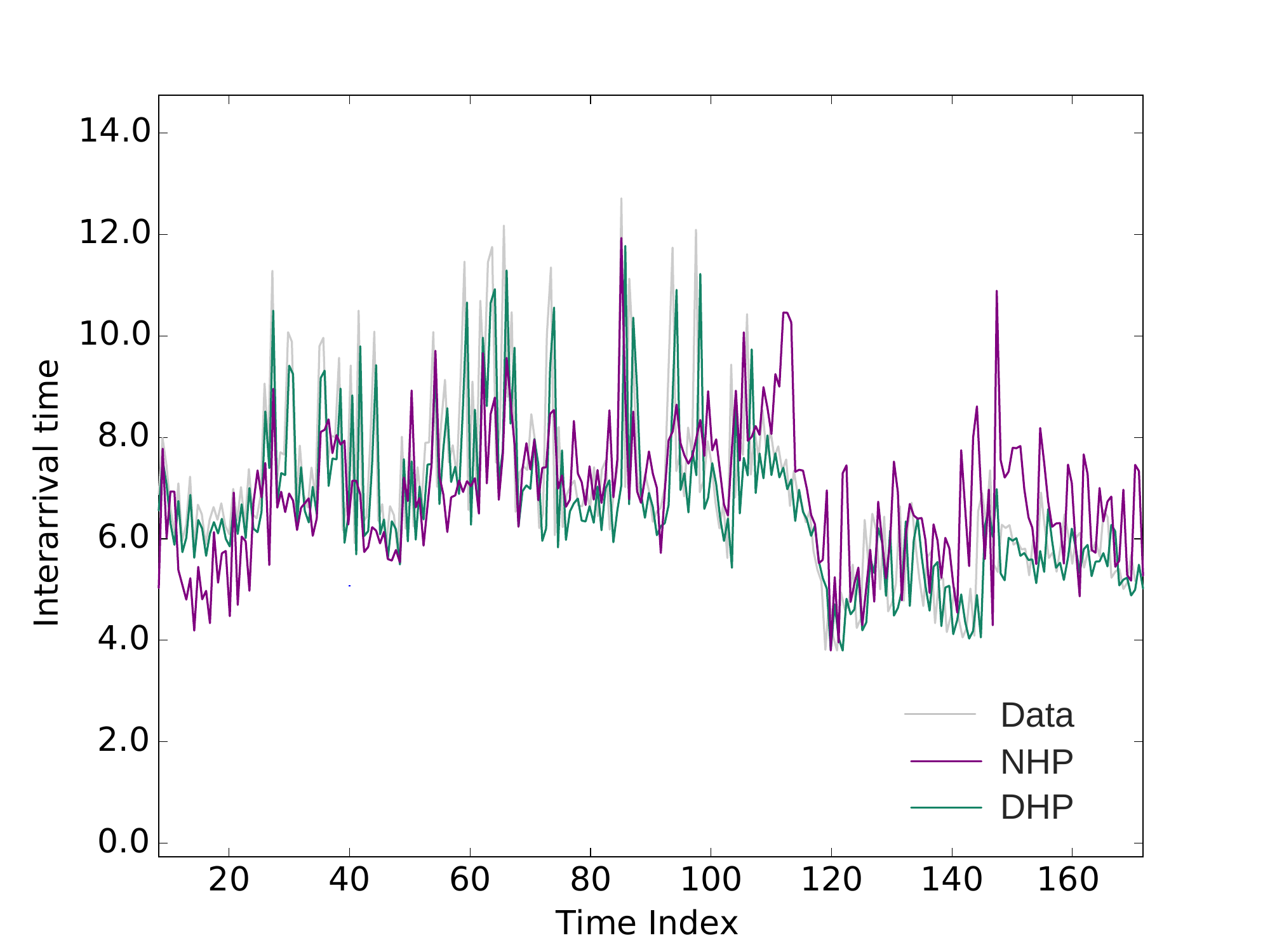}
  \caption{Time Prediction Graph}\label{fig:IETM}
      \end{subfigure}
    \begin{subfigure}[t]{0.32\textwidth}
        \centering
        \includegraphics[width=\textwidth]{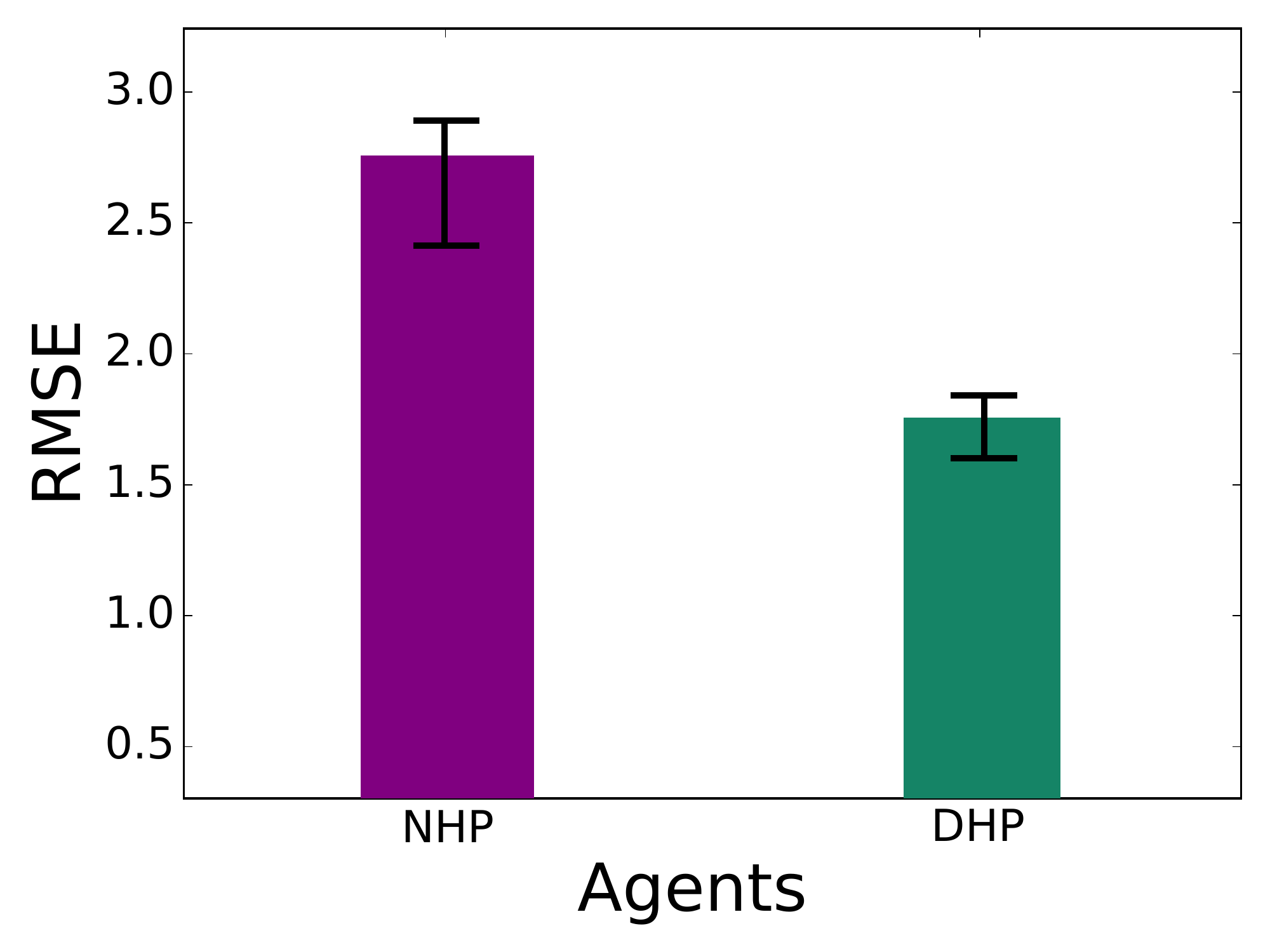}
  \caption{Time Prediction Error}\label{fig:RMSEM}
    \end{subfigure}
    \begin{subfigure}[t]{0.32\textwidth}
        \centering
        \includegraphics[width=\textwidth]{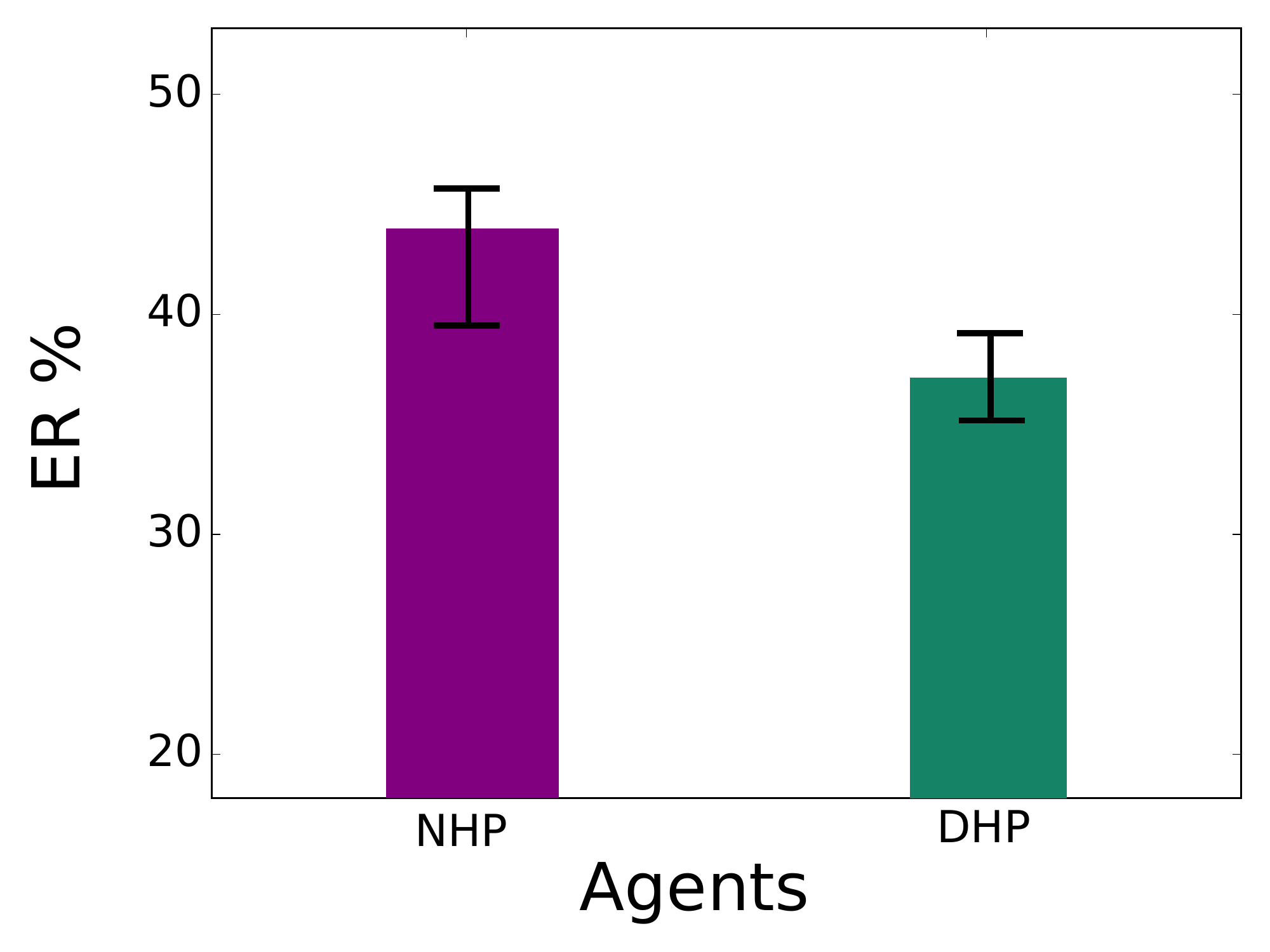}
  \caption{Event Prediction Error}\label{fig:ERM}
      \end{subfigure}
          
   \caption{Performance evaluation of high-frequency market making agents in predicting order book events and time at millisecond resolution. The standard deviation over 10 experiments using different train-val-test sample is denoted by error bar. }\label{fig:PTEM}
 
\end{figure}

In Figure \ref{fig:PTEM}, we evaluate the predictive performance of the high-frequency market making agents using millisecond data sampled from the reconstructed order book. Compared to the earlier results, the DHP model's performance has drastically increased. In addition, the time prediction is consistently better than with the NHP. The deep model with novel architecture and pre-training module are sophisticated enough to capture excitation and feedback effects in the order book. The results also substantiate the claim of the NHP model regarding stochastically missing data. The order book events at millisecond timestamps theoretically omit the events at the finer time resolution, but they are generated by a different mechanisms. This is the reason why there are completely different results at the two time resolutions. The Deep Hawkes model presented here is expressive enough to learn true predictive distribution with scholastically missing events, but only if they are generated from the same mechanism. By integrating the predictive  capabilities into the market making strategies, the agents trade in a simulation framework populated with heterogeneous trading strategies. In the next section, we explore the agents' trading performance.
 
\subsection{Trading Performance}

The trading performance of the high-frequency market making agents is discussed in  Table \ref{tab:DHP_AP}. Our simulation framework evaluates various models, including the Neural Hawkes model, the benchmark probabilistic estimate model, and our proposed Deep Hawkes model. According to the performance metrics specified in Table \ref{tab:DHP_AP}, our proposed agent using DHP (DHMM) consistently outperforms the PMM and NHMM, which suggests that the proposed trading agent benefits by learning the robust microstructure of order book data. Further, the DHP lets the agent capture the self- and cross-excitation effects of the limit order book together with a feedback loop, to place the right order at the right time. We discuss the performance of each agent in detail below.

\begin{figure}[H]
    \centering
    \begin{subfigure}[t]{0.32\textwidth}
        \centering
        \includegraphics[width=\textwidth]{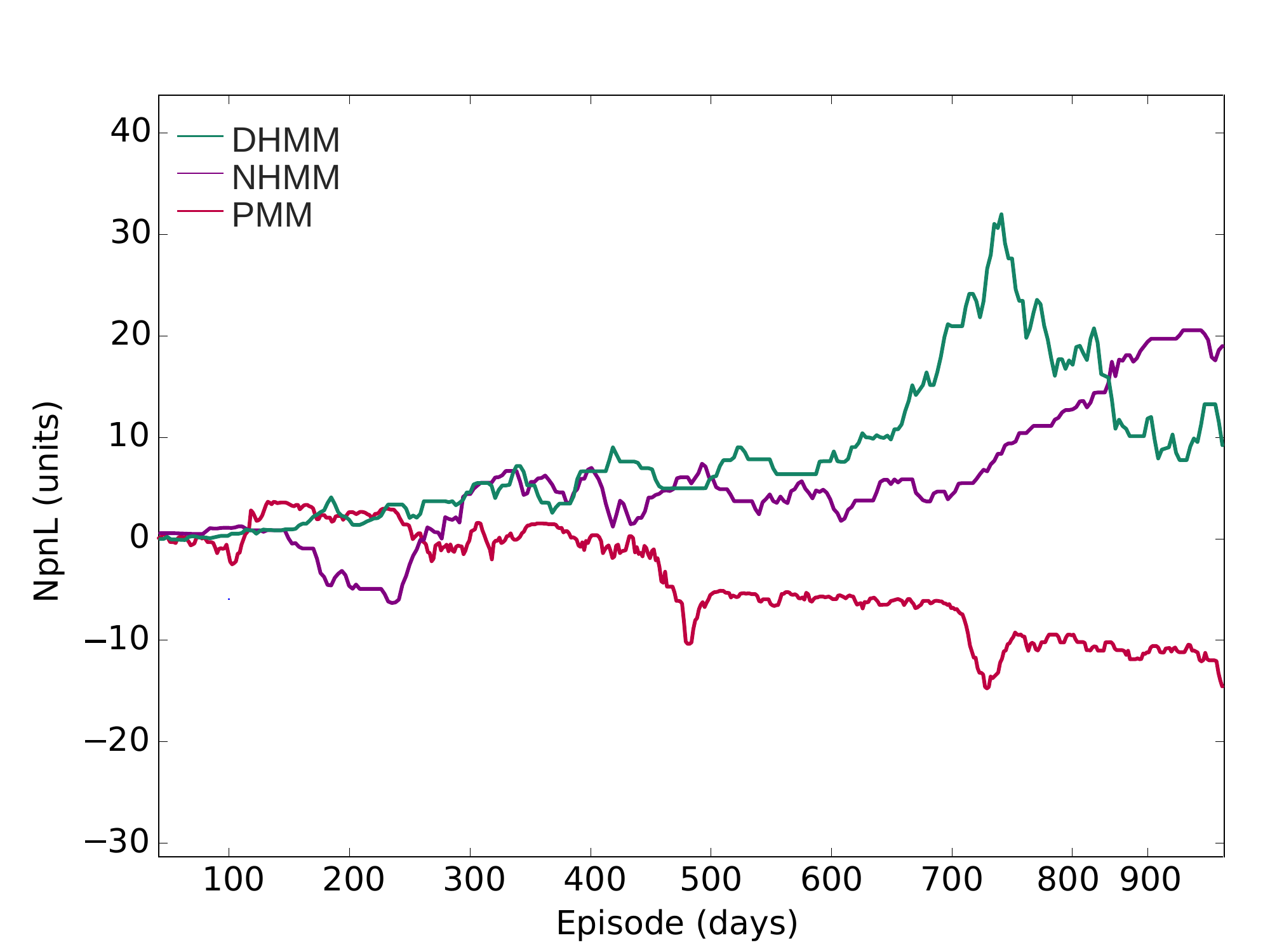}
  \caption{Train}\label{fig:DHP_RTR}
      \end{subfigure}
    \begin{subfigure}[t]{0.32\textwidth}
        \centering
        \includegraphics[width=\textwidth]{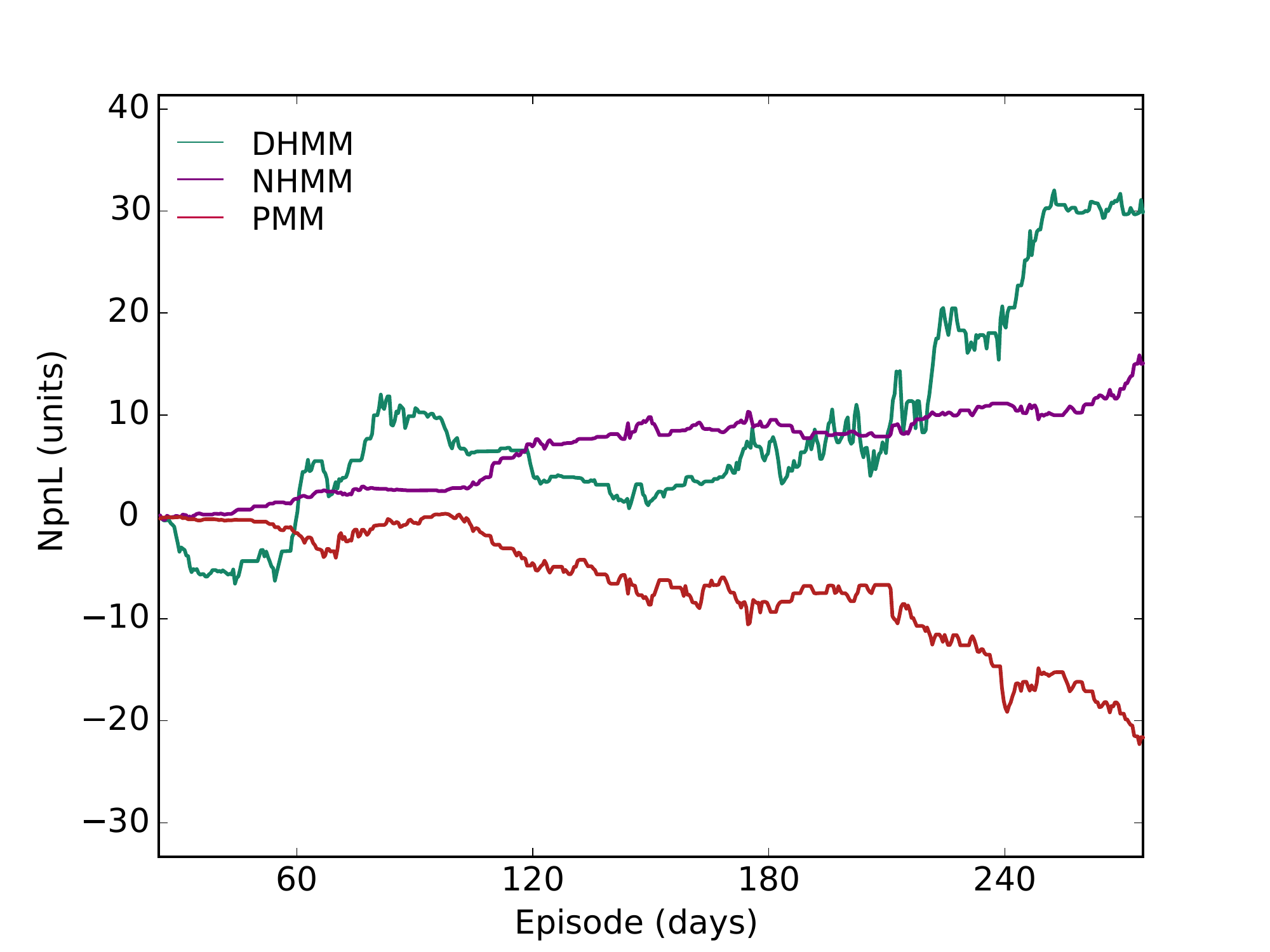}
  \caption{Test}\label{fig:DHP_RTS}
    \end{subfigure}
    \begin{subfigure}[t]{0.32\textwidth}
        \centering
        \includegraphics[width=\textwidth]{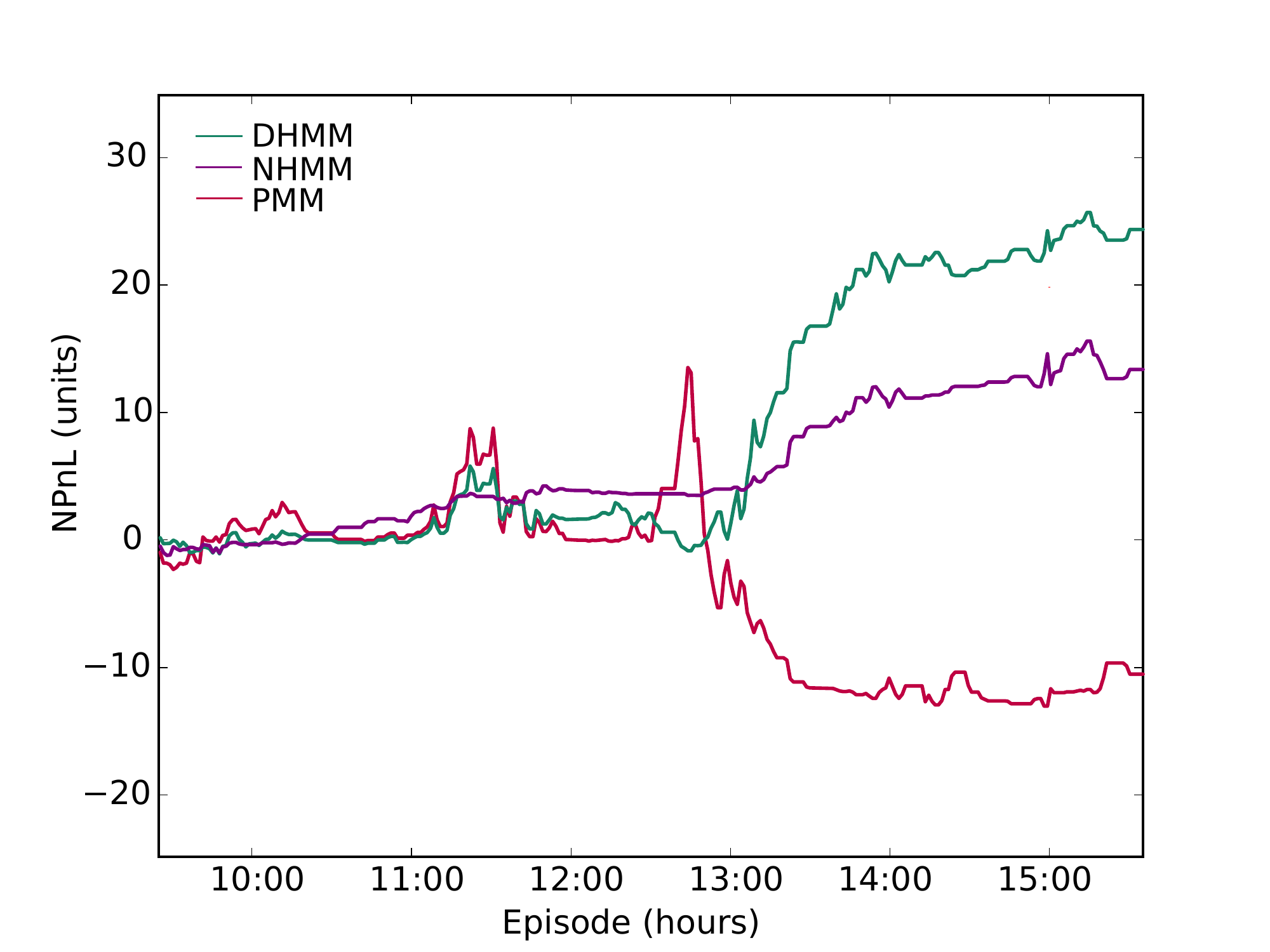}
  \caption{Random day}\label{fig:DHP_RDH}
      \end{subfigure}
          
   \caption{Trading agents performance with DHMM, NHMM and PMM while training, testing and random day. }\label{fig:DHP_TD}
   \end{figure}

As shown in Figure \ref{fig:PTEM}, the DHMM is better at predicting the type of order and its time compared to NHMM. This is an important element
of the market making strategy. The novel DLSTM-SDAE architecture allows the agents to learn the hidden representation of the noisy order book data, and therefore to place orders that add to its profitability. Furthermore, DHMM exhibits a faster convergence rate compared to
NHMM, as shown in Figure \ref{fig:DHP_TD}.

The baseline strategy used by PMM maintains a probability density
estimated on the basis of fundamental price. The fundamental price
evolves according to the jump process, following a normal distribution.
This works in the favor of PMM, which makes more profit at the
beginning, as verified in Figure \ref{fig:DHP_TD}.Over time, however, the DHMM and NHMM learn the art of placing the right order with the right
intensity. Afterwards, the profitability of the PMM fall dramatically.
The PMM might perform better if it took a long position over several
days, rather than trading intraday. It would be interesting to check the
performance of the PMM agents with different probability density
estimate conditions on the joint distribution of microstructure features.  

   \begin{table}[H]
   \centering
\caption{Mean and standard deviation on the daily normalised PnL (PnL) and mean absolute positions (MAP) for different market makers  }\label{tab:DHP_AP}
 \resizebox{\textwidth}{!}{  \begin{tabular}{cllllll}
    \toprule
    \toprule
    \multirow{2}{*}{Agents} & 
      \multicolumn{2}{c}{NPnL [$10^5$]} &
      \multicolumn{2}{c}{MAP[unit]} \\
      \cmidrule(l){2-3} \cmidrule(l){4-5}
    & Mean & Std.Dev. & Mean & Std.Dev. \\
    \midrule
    DHMM & 2.1 & $\pm 18.26$ & 17 & $\pm 20$  \\
    \midrule
    NHMM & 1.1 & $\pm 4.09$ & 4 & $\pm 6$ \\
    \midrule
    PMM & -1.6  & $\pm 79.55$ & 41 & $\pm 74$  \\
   \bottomrule
    \bottomrule
  \end{tabular}
}  
\end{table}
 
\subsection{Order Cancellation Effect} 

Massive numbers of order cancellations in a short period are a
distinctive attribute of the equity market. For example, at Nasdaq
Nordic, order cancellations typically account for 40\% of submitted limit
orders on a particular trading day. Market making strategies using limit
order cancellations contribute to the market marker's profit, bid-ask
spread and order queue position \citep{Dahlstrm2018}.We study the
distribution of profit, bid-ask spread and order queue position by
removing the cancellations mechanism in the base simulation
framework. We estimate the intrinsic value of the order relative to the
queue position by applying the model developed by \cite{Moallemi2017}. The agent's order queue position provides an estimate of number
of orders ahead of the agent’s order at a particular price. A position at
the front of the queue guarantees prompt execution, higher fill rate, low
latency and lower adverse selection cost. We estimate the queue position
in the order book by reconstructing the limit order book from the
simulated data feed.

Lets us suppose that the high-frequency market making agent places a limit order at time $t\!=0$ seeking best ask price $p_a$ which gets filled or canceled at time $\tau$. Filling the order pays the agent $p_a$ while cancellations pay nothing. We now describe the value of the order perceived by agents relative to the queue position as:

\begin{equation}
\begin{split}
V_t &= \mathbb{E}[  (p_a - p_t)\mathbb{I}_{\mathsf{FILL}} -  (p - p_t)\mathbb{I}_{\mathsf{FILL}}  \mid \mathcal{F}_t ] \\
&= \mathsf{FP}_t (\mathsf{LSP}_t - \mathsf{ASC}_t) \\
&= fill \; probablity \left( liquidity \; spread \; premium - adverse \; selection \; cost \right)
\end{split}
\label{EQ:COE}
\end{equation} 

where 
\begin{eqnarray*}
\begin{aligned}
\!& \mathsf{FP}_t  \triangleq  \mathbb{P} \left( \mathsf{FILL} \mid \mathcal{F}_t \right), \\
\!& \mathsf{LSP}_t  \triangleq  \left( p_a - p_t \right), \\
\!& \mathsf{ASC}_t  \triangleq  \mathbb{E} \left[ (p_{\tau} - p_t) \mid \mathcal{F}_t, \mathsf{FILL} \right]
\end{aligned}
\end{eqnarray*} 

To empirically calibrate the model (Equation \ref{EQ:COE}), we take the same parameters used by \cite{Moallemi2017}. These are exponential order size distribution, trade arrival rate (TAR), average trades size (ATS), trade size in the stan-dard lot (TSS), cancellation arrival rate (CAR), average cancellation size (ACS), price jump arrival rate (PJR), average jump size (AJS), market impact (MI) and average queue size (AQS). The trade size is identified as the limit order or market order, contrary to aggressive market orders as described by \cite{Moallemi2017}. Table \ref{tab:OCE} specifies the estimated parameters for simulated data with no cancellation mechanisms (Simulated NC), without cancellation mechanisms (Simulated WC) and an average (Simulated AV) over 21 days. The paper itself provides more detail regarding the parameters, calibration and model fitting \citep{Moallemi2017}.

\begin{table}[H]
\caption{Estimated parameters for simulated orderbook data.}\label{tab:OCE}
\resizebox{\textwidth}{!}{  \begin{tabular}{cccccccccc}
    \toprule
    \toprule
    
     Data & TAR (/min)& ATS (shares) & TSS (shares) & CAR (/min) & ACS (shares) & PJR (/min) & AJS (ticks) & MI   & AQS (shares) \\ [0.5ex]
    \midrule
    Simulated NC & 2.53 & 3467 &7664 & 92.72  & 5061 & 1.26 & 0.32 & 10.91 & 16416  \\
    Simulated AV & 2.04 & 4037 & 6901 & 82.21 & 4022 & 1.01 & 0.46 & 8.76 & 23554  \\
    \midrule
    Simulated WC & 5.26 & 6329 & 8083 & 43.71 & 1107 & 3.75 & 2.06 & 11.92 & 40191\\
     Simulated AV & 4.07 & 5463 & 9147 & 40.31 & 1560 & 3.01 & 2.06 & 13.02 & 46815 \\
   \bottomrule
    \bottomrule
  \end{tabular}}
  
\end{table}

\begin{figure}[H]
    \centering
    \begin{subfigure}[t]{0.32\textwidth}
        \centering
        \includegraphics[width=\textwidth]{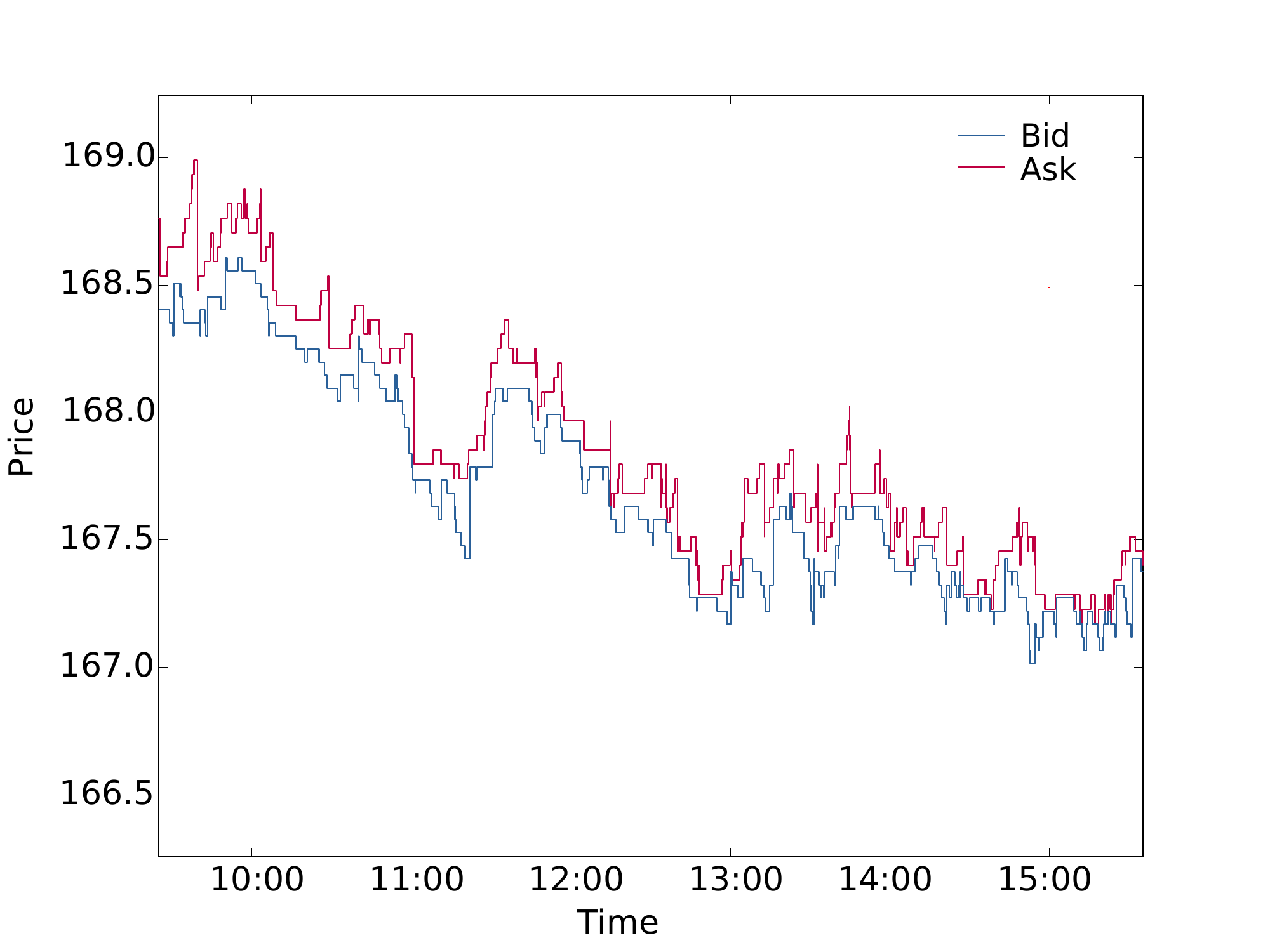}
  \caption{BAD WC}\label{fig:BAD_B}
      \end{subfigure}
    \begin{subfigure}[t]{0.32\textwidth}
        \centering
        \includegraphics[width=\textwidth]{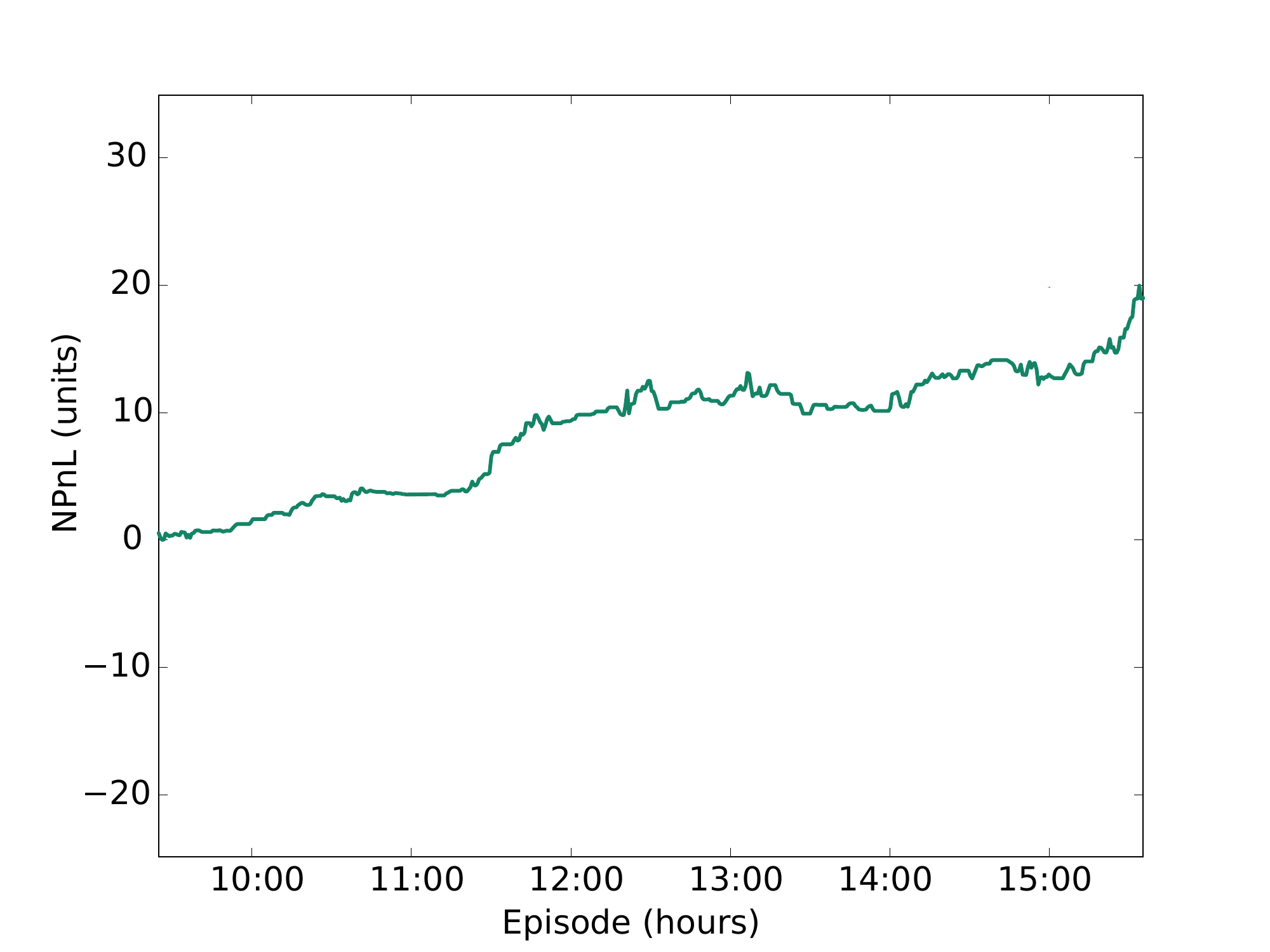}
  \caption{DHMMP WC}\label{fig:RDH_B}
    \end{subfigure}
    \begin{subfigure}[t]{0.32\textwidth}
        \centering
        \includegraphics[width=\textwidth]{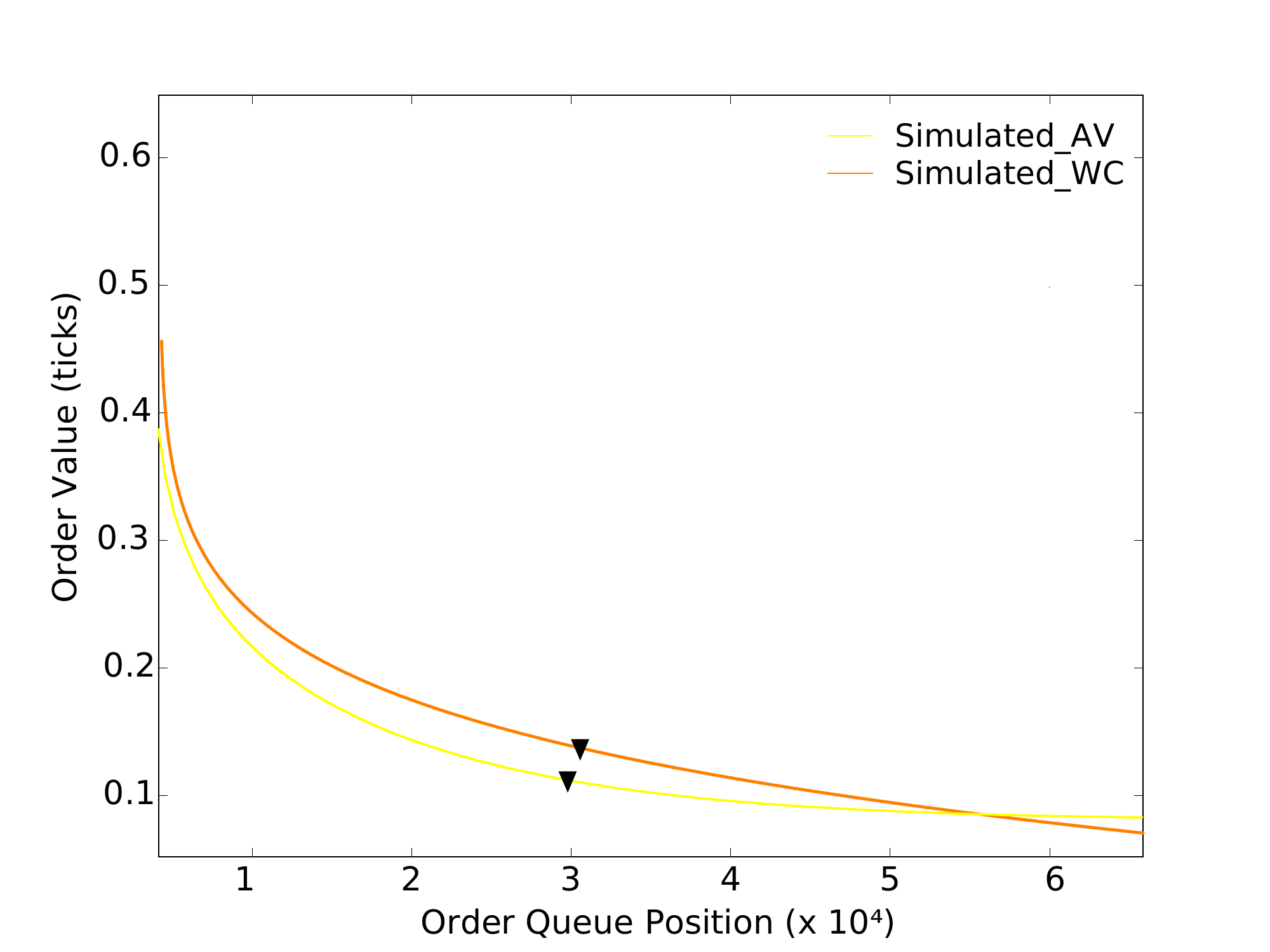}
  \caption{VOPQ WC}\label{fig:VO_B}
      \end{subfigure}
          
      \begin{subfigure}[t]{0.32\textwidth}
        \centering
        \includegraphics[width=\textwidth]{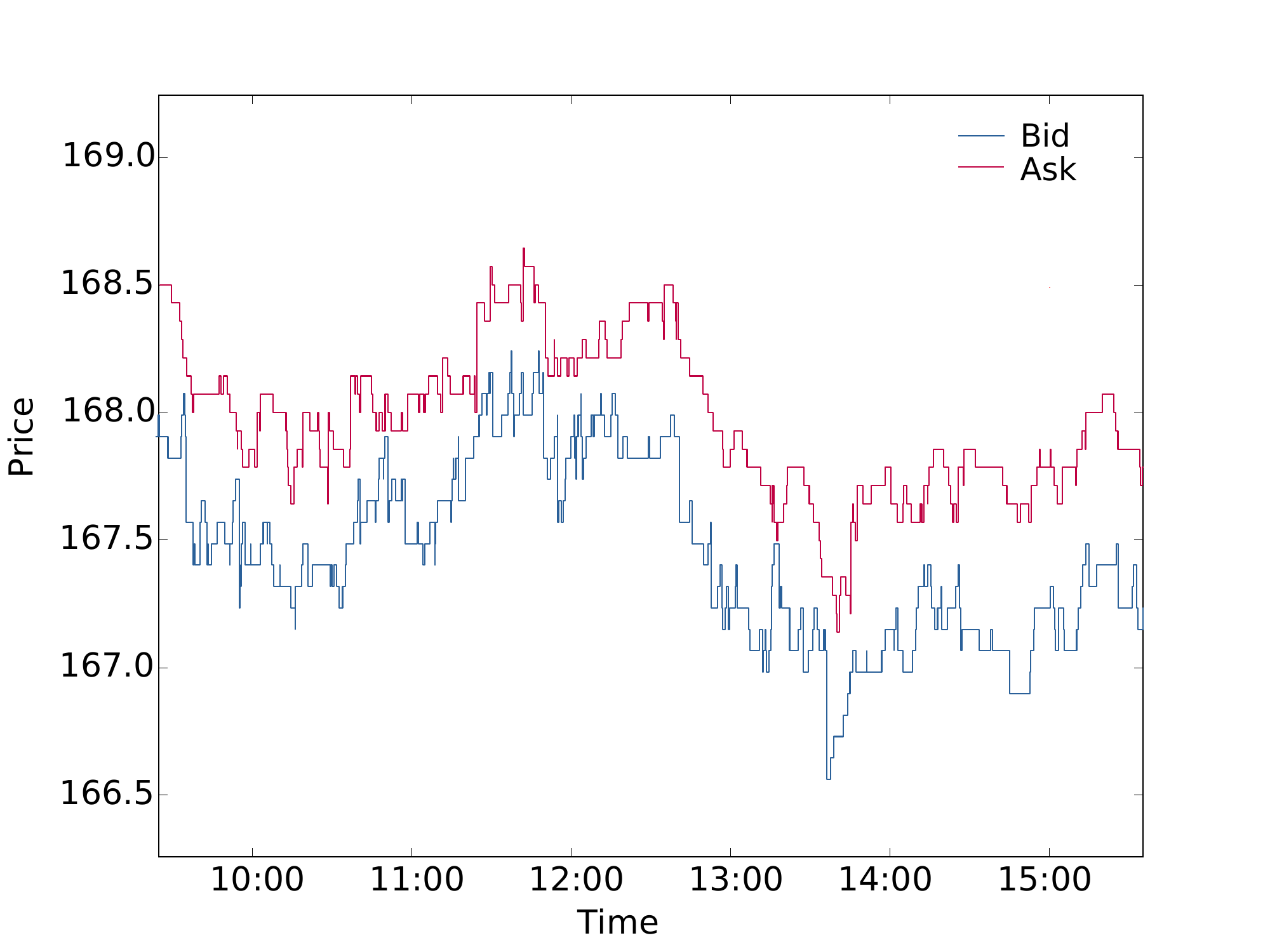}
  \caption{BAD NC}\label{fig:BAD_A}
      \end{subfigure}
    \begin{subfigure}[t]{0.32\textwidth}
        \centering
        \includegraphics[width=\textwidth]{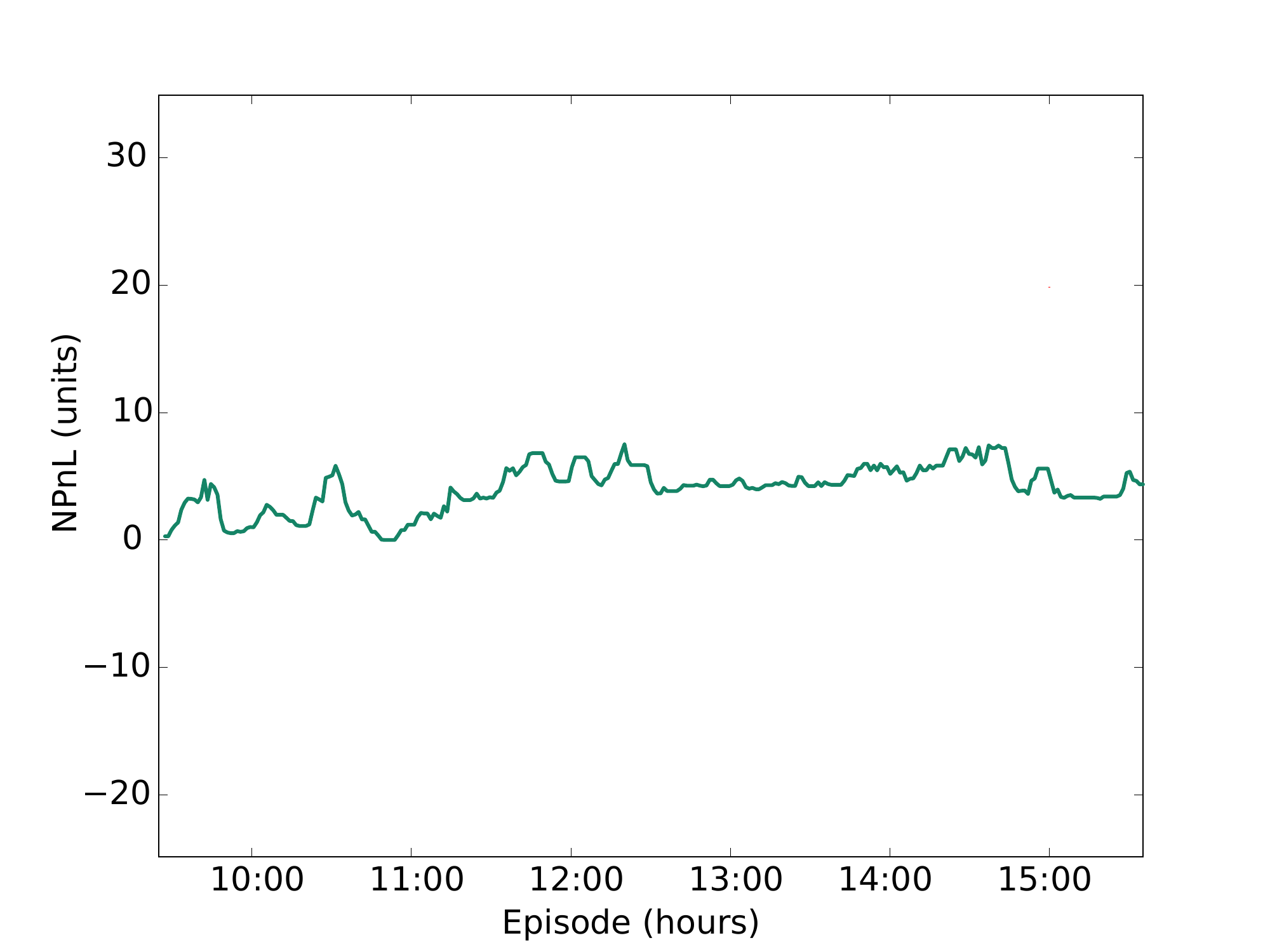}
  \caption{DHMMP NC }\label{fig:RDH_A}
    \end{subfigure}
    \begin{subfigure}[t]{0.32\textwidth}
        \centering
        \includegraphics[width=\textwidth]{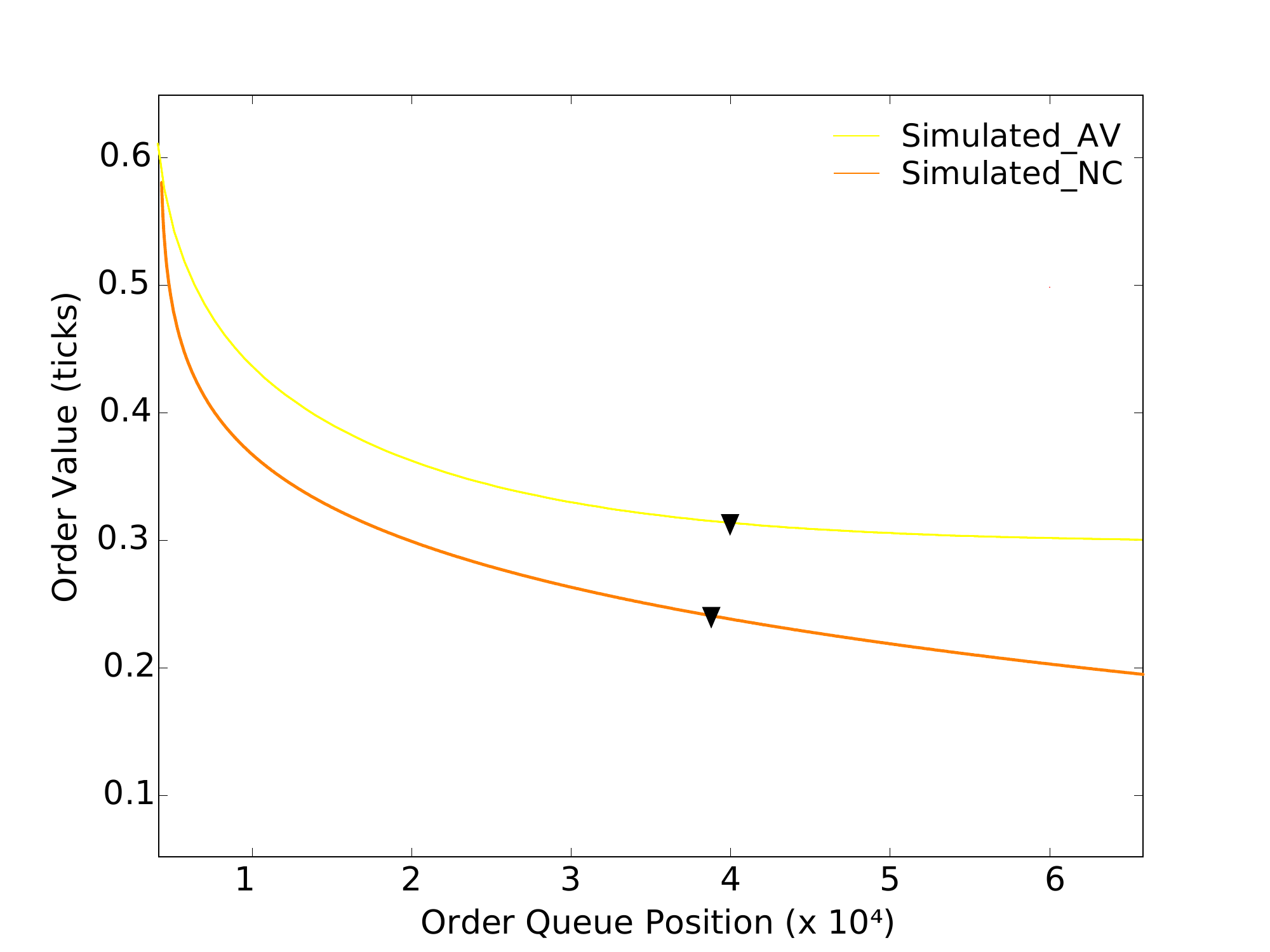}
  \caption{VOPQ NC}\label{fig:VO_A}
      \end{subfigure}
          
   \caption{Effect of limit order cancellations on the market. The top
row (marked WC) represents the distribution when the market maker’s
agents can cancel the limit orders. The bottom row (marked NC)
represents a situation with no cancellations. BAD is intraday bid-ask
distribution, DHMMP is profit distribution of DHMM over the trading
day, and VOPQ is the value of the orders relative to queue position. The
average queue length on a particular trading day is represented by a
black triangle.}\label{fig:OCE}
   \end{figure}

Table \ref{tab:OCE} shows that an absence of cancellation mechanisms at the high-frequency market maker's end leads to a drastic increase in the average queue size. The decrease in the cancellation rate increases the queue size, which affects the high-frequency market maker's profitability, bid-ask spread and market impact. The value of the order as a function of the queue position, bid-ask spread, and the agent's profit efficiently captures the claim illustrated by the data in Figure \ref{fig:OCE}. The wider bid-ask spread when agent's are unable to cancel the limit order in Figure \ref{fig:BAD_A} has negative effect on the profitability (Figure \ref{fig:RDH_A} ) as compared to scenarios with cancellations (Figure \ref{fig:BAD_B},\ref{fig:RDH_B} ). As stated in the model in Equation \ref{EQ:COE}, the value of an order that is not filled is zero. Figure \ref{fig:VO_A} shows that an increase in queue length decreases the probability of execution, and therefore the value. The value of the order becomes flat, as the queue length is extremely large. Our results are consistent with the findings of \cite{Dahlstrm2018} when investigating the determinants of order cancellations. It is difficult to infer causal relationships between cancellations and market microstructure variables based on artificially created scenarios, but this approach nonetheless it paves the way for future investigation using order level data.

\subsection{Sensitivity Analysis}
We perform sensitivity analysis on the proposed model by changing the
hyperparameters – specifically, the number of layers, the number of
hidden units per layer and the noise level. The aim is to confirm the
robustness of the model, rather than overfitting parameters. Table \ref{tab:SA} shows the performance of our proposed model when there is an increase in the number of layers, hidden units and noise level. The model is robust to the noise level at the optimal choice of numbers of layers, parameters and hidden units.  

\begin{table}[H]
\caption{Sensitivity to the number of hidden units and Gaussian noise.
The DLSTM-SDAE used in our model has 3 DAE layers and 3 LSTM layers. In performing sensitivity analysis, we fix the 3 LSTM layers and
change only the DAE layer. }\label{tab:SA}
 \resizebox{\textwidth}{!}{ \begin{tabular}{llll}
    \toprule
    \toprule
    & \# Number of layer 1 & \# Number of layer 2 & \# Number of layer 3 \\
    \midrule 
    Number of hidden unit per layer & $\left(64, 128, 256, 512, 1024 \right)$ & $\left(64, 128, 256, 512, 1024 \right)$ & $\left(64, 128, 256, 512, 1024 \right)$ \\
    
    \midrule
    RMSE &  $\left( 4.6, 4.4, 4.1, 4.1, 4.0 \right)$ & $\left( 3.5, 3.0, 3.0, 2.5, 2.5 \right)$ & $\left( 2.0, 2.0, 2.0, 1.5, 1.5 \right)$ \\
    \midrule
    ER & $\left( 55.7, 55.7, 55.7, 50.4, 54.0 \right)$ & $\left( 47.2, 44.4, 44.1, 44.1, 44.1 \right)$ & $\left( 37.5, 37.0, 35.0, 35.0, 34.0 \right)$ \\
   \midrule
   Gaussian noise & $\left(0.10, 0.20, 0.30, 0.40, 0.50 \right)$ & $\left(0.10, 0.20, 0.30, 0.40, 0.50 \right)$& $\left(0.10, 0.20, 0.30, 0.40, 0.50 \right)$\\
   \midrule
    RMSE & $\left(7.2,5.9, 4.3, 5.4, 6.5 \right)$& $\left(4.1,3.9, 3.0, 4.0, 4.6 \right)$&$\left(2.2,2.2, 2.0, 2.1, 2.1 \right)$ \\
    \midrule
    ER & $\left(60.2,55.9, 50.1, 60.0, 63.5 \right)$ & $\left(55.6,52.2, 49.1, 55.3, 67.5 \right)$& $\left(37.2,37.9, 37.3, 37.1, 37.2 \right)$ \\
    
     \bottomrule
    \bottomrule
  \end{tabular}}
  
\end{table}

\subsection{Validation}

\begin{figure}[H]
    \centering
    \begin{subfigure}[t]{0.32\textwidth}
        \centering
        \includegraphics[width=\textwidth]{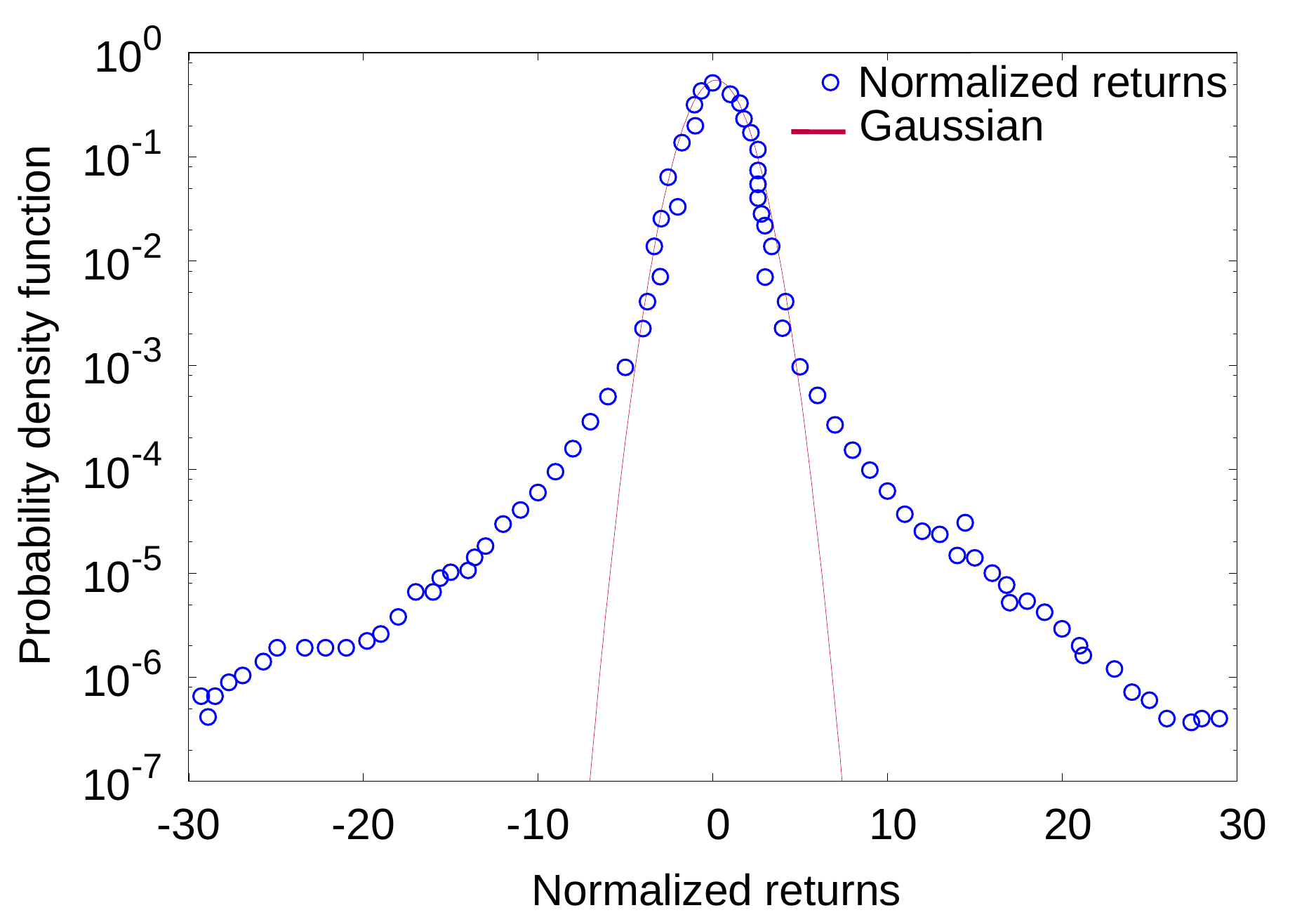}
  \caption{PDF}\label{fig:DHP_N}
      \end{subfigure}
    \begin{subfigure}[t]{0.32\textwidth}
        \centering
        \includegraphics[width=\textwidth]{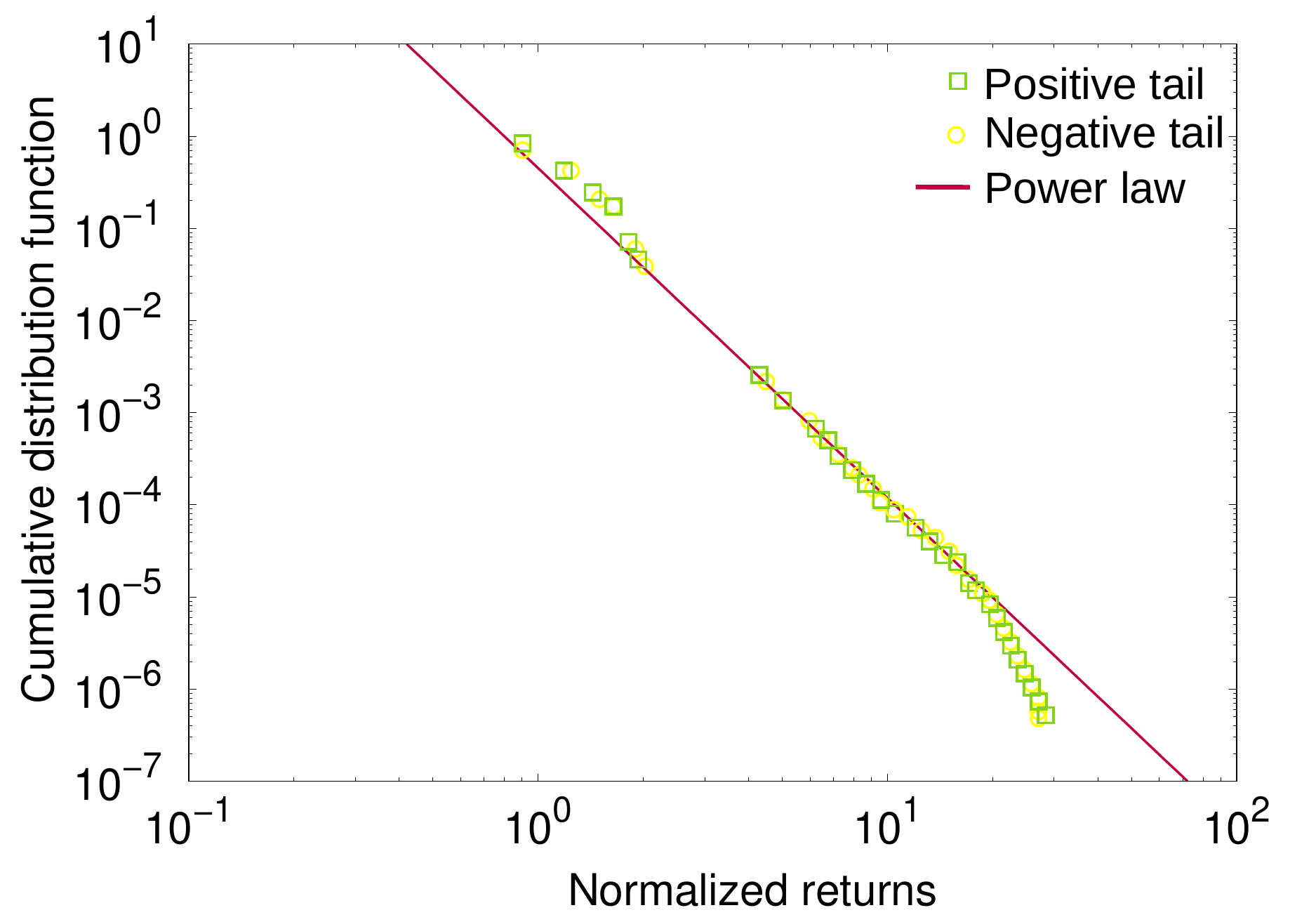}
  \caption{CDF}\label{fig:DHP_NCDF}
    \end{subfigure}
    \begin{subfigure}[t]{0.32\textwidth}
        \centering
        \includegraphics[width=\textwidth]{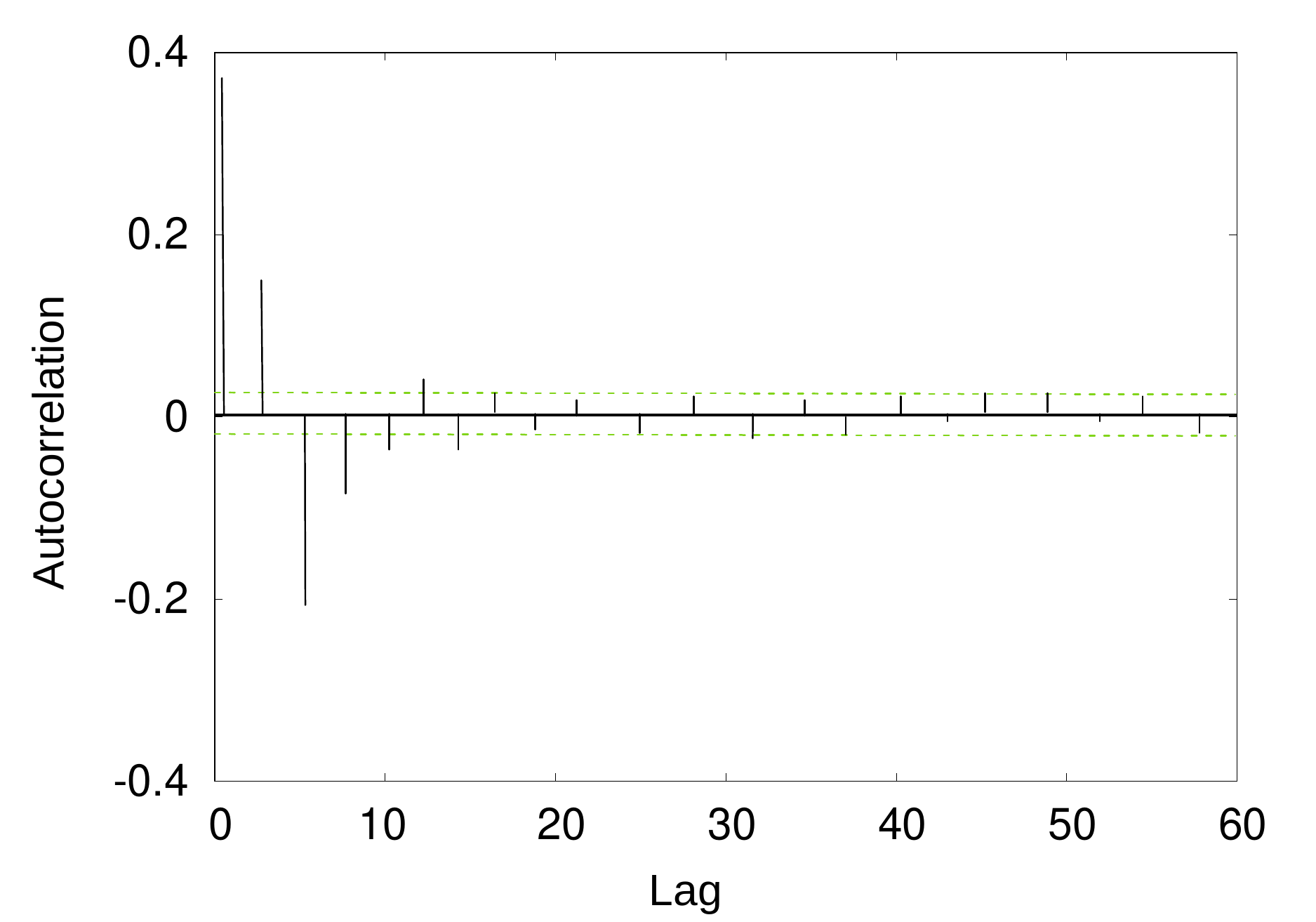}
  \caption{ACF}\label{fig:DHP_SFAC}
      \end{subfigure}
          
   \caption{Stylised facts reproduced from simulated order book data. All graphs were generated on the basis of $\Delta t = 1\; minute $. }\label{fig:DHP_SF}
   \end{figure}

The validation of the trading simulation framework is performed by
measuring how successfully the simulation's output exhibits persis- tent
empirical patterns in the order book data. Such empirical patterns are
common across various markets and instruments, and even timescales
are often classified as "stylised facts" \citep{Cont2001}. We present a
nominal set of stylised facts, reproduced from the empirical analysis of
simulated order book data, as shown in Figure \ref{fig:DHP_SF}.

Let $p(t)$ be the price of a security at time $t$. Given a timescale $\Delta t$, we define log return at $\Delta t $ as $ r (t, \Delta t) = \ln p(t+\Delta t) - \ln p(t)$. The cumulative distribution (CDF) of returns is given as $F_{\Delta t}(x) = \mathbb{P}[r(t, \Delta t) \leq x]$. The derivative of the earlier gives probability density function (PDF) $F^{'}_{\Delta t}=f_{\Delta t}$, empirically estimated for normalized simulated return, as illustrated in Figure \ref{fig:DHP_N}. The cumulative distribution of return follows power law $F_{\Delta t} \sim \left|r\right|^{-\alpha }$ with $2<\alpha<5$. In Figure \ref{fig:DHP_NCDF},  the positive tail $F^{+}_{\Delta t}(x) = \mathbb{P}[r(t,\Delta t) \geq x]$ and the negative tail $F^{-}_{\Delta t}(x) = \mathbb{P}[r(t,\Delta t) \leq x]$ of cumulative distribution, shown as yellow circles and green squares, exhibit power law, as denoted by the red line with $\alpha=2.8$. In Figure \ref{fig:DHP_SFAC}, we show the absence of the autocorrelation  of price change, defined as $\rho (\tau) = Corr \left( r (t, \Delta t ), r (t + \tau, \Delta t )  \right) $.  The autocorrelation function (ACF) drastically decays to zero in few lags. 

%
\section{Conclusions}
\label{sec:DHP_C}

We have developed a market making strategy that takes account of the
feedback loop between the order arrival and the state of the LOB, with
self- and cross-excitation effects, while placing an order in our realistic simulation framework. The strategy was designed by integrating the self-modulating multivariate Hawkes process with DLSTM-SDAE. The
data-driven approach performed adversely in relation to predicting the
next order type and its timestamp when fitted to reconstructed order
book data at nanosecond resolution. When trained with millisecond
resolution data, it outperforms NHP in prediction tasks and benchmark
market making strategies in trading performance. We have demonstrated that extending the DHP in a market making setting accomplished better performance when validating empirical claims about the effect of cancellation on the determinants of order size. Our modelling approach is still far from inferring causality, but does pave the way exploring a range of diverse research avenues. The most important and immediate of these are listed below:
\begin{enumerate}
\item Apply more advanced pre-processing, architecture, and learning
algorithms to filter out ultra-noisy order book data at nanosecond
timestamps.
\item Explore an intensity-free approach for Hawkes processes with learning mechanisms other than the maximum likelihood approach.
\item Embedding DHP within the reinforcement learning framework to learn
the optimal policy.
\item Extend the model to the deep reinforcement learning framework, in
which the agent’s trading action and reward from the simulator are
asynchronous stochastic events characterised by marked multivariate
Hawkes processes. 
\item Extract the agent’s trading algorithm parameters directly from the order book data , rather than from random seeds or empirical literature.
\end{enumerate}


\newpage
\bibliographystyle{apalike} 
\bibliography{ref/ref}

\end{document}